\documentclass[iop]{emulateapj}
  \usepackage{natbib}
\newcommand{\teff}{T$_{\rm eff}$}
\newcommand{\kms}{km~s$^{-1}$}
\newcommand{\vt}{v$_{t}$}

\newcommand{\rgc}{R$_{\rm gc}$}

\slugcomment{Published in The Astronomical Journal, 2013, 145, 107}

\shorttitle{Open Cluster Neutron Capture Abundances}
\shortauthors{Jacobson \& Friel.}

\defcitealias{2012AJ....144...95Y}{Y12}
\defcitealias{2011ApJ...736..120M}{M11}
\defcitealias{2009ApJ...693L..31D}{D09}

\begin{document}

\title{Zirconium, Barium, Lanthanum and Europium Abundances in Open Clusters}

\author{Heather R. Jacobson\altaffilmark{1,}\altaffilmark{2,}\altaffilmark{3}}
\affil{Department of Physics \& Astronomy, Michigan State University, East Lansing, MI 48823;
jacob189@msu.edu}

\author{Eileen D. Friel}
\affil{Department of Astronomy, Indiana University, Bloomington, IN 47405;
efriel@indiana.edu}

\altaffiltext{1}{National Science Foundation Astronomy and Astrophysics Postdoctoral Fellow.}
\altaffiltext{2}{Visiting Astronomer, Kitt Peak National Observatory, National Optical Astronomy Observatory, which is operated by the Association of Universities for Research in Astronomy (AURA) under cooperative agreement with the National Science Foundation.}
\altaffiltext{3}{Current address: Massachusetts Institute of Technology, Kavli Institute for Astrophysics and Space Research, 77 Massachusetts Ave., Cambridge, MA, 02139; hrj@mit.edu}

\begin{abstract}
We present an analysis of the s-process elements Zr, Ba, and La and the r-process element Eu in a sample of 50 stars in 19 open clusters.  Stellar abundances of each element are based on measures of a minimum of two lines per species via both equivalent width and spectrum synthesis techniques.  We investigate cluster mean neutron capture abundance trends as a function of cluster age and location in the Milky Way disk and compare them to results found in other studies in the literature.  We find a statistically significant trend of increasing cluster [Ba/Fe] as a function of decreasing cluster age, in agreement with recent findings for other open cluster samples, supporting the increased importance of low mass AGB stars to the generation of s-process elements.  However, the other s-process elements, [La/Fe] and [Zr/Fe] do not show similar dependences, in contrast to theoretical expectations and the limited observational data from other studies.  Conversely, cluster [Eu/Fe] ratios show a slight increase with increasing cluster age, although with marginal statistical significance.  Ratios of [s/r]-process abundances, [Ba/Eu] and [La/Eu], however, show more clearly the increasing efficiency of s-process relative to r-process enrichment in open cluster chemical evolution, with significant increases among younger clusters.  Lastly, cluster neutron capture element abundances appear to be independent of Galactocentric distance.  We conclude that a homogeneous analysis of a larger sample of open clusters is needed to resolve the apparent discrepant conclusions between different studies regarding s-process element abundance trends with age to better inform models of galactic chemical evolution.

\end{abstract}


\section{Introduction}
It is well established that open clusters do not exhibit an age-metallicity relation (e.g., \citealt{1995ARA&A..33..381F}, \citealt{2011A&A...535A..30C}).  The fact that old, super-solar metallicity clusters (e.g., NGC 188, NGC 6791) and young, sub-solar metallicity clusters (e.g., NGC 2324) can be found in the disk seems to indicate that the location of a cluster's birth is more important than when it formed in terms of the signatures of disk chemical evolution imprinted on its stars \citep{2004A&A...414..163S}.  This is seen in the radial metallicity gradient in the disk, as traced by open clusters, cepheids, and other populations (e.g., \citealt{2012AJ....144...95Y}).  The  distribution of abundances of other chemical elements (such as the alpha-, light and Fe-peak elements) also appear to trace that of iron very well, with open clusters having [X/Fe] ratios that are essentially independent of cluster age and galactocentric radius (\rgc; e.g., \citealt{2009A&A...494...95M}, \citealt{2010AJ....139.1942F})\footnote{This discussion ignores the possible role of radial migration in shaping trends of cluster abundances (and age) with \rgc.  While recent simulations have indicated that radial migration of stars in galaxy disks via various mechanisms can greatly influence radial metallicity gradients (e.g., \citet{2008ApJ...684L..79R} and \citet{2011ApJ...737....8L}), to our knowledge it has not been demonstrated that open clusters are subject to radial migration.  It goes without saying the usefulness of cluster element abundance trends with \rgc\ as probes of disk chemical evolution hinges on the answer to this question.}.

The findings of studies of open cluster neutron-capture element abundances, however, stand in exciting contrast to the behavior seen for the other elements.  Neutron capture species, generated in low mass AGB stars and massive stars via the strong and weak s-processes and (likely) in type II supernovae via the r-process (e.g., \citealt{1999ARA&A..37..239B}, \citealt{2001ApJ...557..802B}, \citealt{2008ARA&A..46..241S}, \citealt{2010ApJ...710.1557P}), have only recently been studied in decently large, homogeneous samples of open clusters, and are further useful probes of the mass range of stars that contributed to the chemical evolution of the disk.  \citet{2009ApJ...693L..31D} reported an unexpected trend of increasing cluster [Ba/Fe] with decreasing cluster age that cannot be explained by chemical evolution models using standard AGB star s-process yields.  An analysis of the s-process elements Y, Ce, Zr and La in open clusters by \citet{2011ApJ...736..120M} found an analogous trend for these elements, though it should be noted that the result is most robust for Y and Ce.  Given that this trend is seen for both light s-elements (Y, Zr) and heavy s-process elements (Ba, La, Ce) in these studies, one can conclude that the cause of this increased enrichment in younger stars affects all elements produced by the main s-process in the same way.

 \begin{deluxetable*}{lccclcl}
\tabletypesize{\footnotesize}
\tablewidth{0pt}
\tablecaption{Cluster Sample\label{cluster_sample}}
\tablehead{
\colhead{} & \colhead{Age\tablenotemark{a}} & \colhead{{\it d}} & \colhead{R$_{gc}$\tablenotemark{b}}   & \colhead{Ref.\tablenotemark{c}} & \colhead{No. of Stars} & \colhead{Telescope}\\ 
\colhead{Cluster} & \colhead{(Gyr)} & \colhead{(kpc)} & \colhead{(kpc)} & \colhead{} & \colhead{} &
\colhead{} }
\startdata
Be 17 & 10.1 & 2.7 & 11.7  & F05 & 3 & KPNO 4m \\
Be 18 & 5.7 & 5.4 & 13.7 & Y12 & 2 & Keck \\
Be 21 & 2.2 & 6.2 & 14.7 & Y12 & 2 & Keck \\
Be 22 & 4.3 & 6.2 & 14.4 & Y12 & 2 & Keck \\
Be 32 & 5.9 & 3.2  & 11.6 & FJP10 & 2, 2 & KPNO 4m, Keck\\
Be 39 & 7.0 & 4.3 & 11.9 & FJP10  & 4 & KPNO 4m\\
M 67 & 4.30 & 0.8 & 9.1 & FJP10 & 3 &  KPNO 4m\\
NGC 188 & 6.3 & 1.7 & 9.4 & FJP10  & 4 & KPNO 4m\\
NGC 1193 & 4.2 & 5.8 & 13.6 & FJP10 &  1 &  HET\\
NGC 1245 & 1.1 & 3.0 & 11.1 & J11 & 4 & KPNO 4m\\
NGC 1817 & 1.1 & 1.5 & 10.0 & J09 & 2 & KPNO 4m\\
NGC 1883\tablenotemark{d} & 0.7 & 3.9 & 12.3 & J09 & 2 & KPNO 4m\\
NGC 2141 & 2.4 & 3.9  & 12.2 & J09 & 2 & KPNO 4m\\
NGC 2158 & 1.9 & 4.0 & 12.8 & J09 & 1 & KPNO 4m\\
NGC 2194 & 0.9 & 1.9 & 10.3 & J11 & 2 & KPNO 4m\\
NGC 2355 & 0.8 & 1.9 & 10.3 & J11 & 3 & KPNO 4m\\
NGC 6939 & 1.2 & 1.8 & 8.4 & A04 & 4 & KPNO 4m\\
NGC 7142 & 4.0 & 1.9 & 9.2 & J08 & 4 & KPNO 4m\\
PWM 4 & 7.0 & 7.2 & 13.3 & Y12 & 1 & Keck\\
\enddata
\tablenotetext{a}{Adopted from \citet{2004A&A...414..163S}}
\tablenotetext{b}{R$_{\odot}$ = 8.5 kpc}
\tablenotetext{c}{References for distances: A04: \citet{2004MNRAS.348..297A}; F05 = \citet{2005AJ....129.2725F}; FJP = \citet{2010AJ....139.1942F}; 
J08 = \citet{2008AJ....135.2341J}; J09 = \citet{2009AJ....137.4753J}; J11 = \citet{2011AJ....142...59J}; Y12 = \citet{2012AJ....144...95Y}}
\tablenotetext{d}{NGC 1883 is not included in the sample of \citeauthor{2004A&A...414..163S}.  This value is taken from \citet{2007MNRAS.379.1089V} (uncertainty 0.07 Gyr).}
\end{deluxetable*}

D'Orazi et al.\ interpreted this anticorrelation with age as an indication that low-mass ($<$1.5 M$_{\odot}$) AGB stars contributed more to the chemical evolution of the disk than theoretical models previously indicated.  This scenario has been further explored by \citet{2012ApJ...747...53M}, who presented revised s-process production models  for low-mass stars in which the neutron source was larger by a factor of four than in previous models.  Applying the s-process yields produced by this model in a chemical evolution study, they could reasonably reproduce the observed trend of [s/Fe] with age seen by open clusters.

\begin{figure*}
\epsscale{0.8}
\plotone{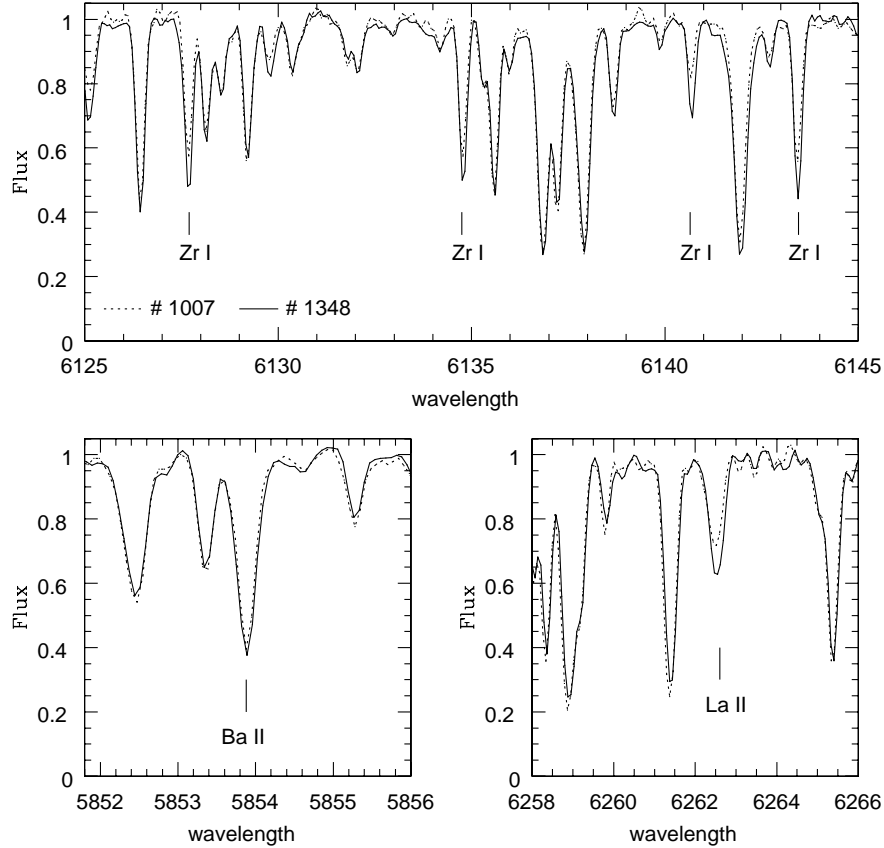}
\caption{Sample portions of the spectra for stars 1007 (dashed line) and 1348 (solid line) of NGC 2141.  The two stars have identical atmospheric parameters and metallicities, but 1348 shows enhanced Zr and La abundances relative to 1007, although their Ba and Eu line strengths are very similar. }
\label{n2141_spec}
\end{figure*}

It is very intriguing that a recent study of solar-type stars in the solar neighborhood (for which there are Hipparcos distances and therefore, isochrone ages) has also found trends of increasing [s/Fe] with decreasing age.  Indeed, the dwarf stars in the study by \citet{2012A&A...542A..84D} show a very similar pattern of [s/Fe] with age to the open clusters in the D'Orazi et al.\ and Maiorca et al.\ studies, most notably for the elements Ba and Zr, and especially for stars younger than the age of the Sun (see their Figure 11).  That said, the results of another open cluster study show conflicting results.  \citet{2012AJ....144...95Y} presented a homogeneous chemical abundance study of 11 open clusters, and in combination of results from the literature, investigated trends of cluster element abundances as a function of \rgc\ and cluster age\footnote{It should be noted for the present discussion that  only a subset of the literature sample they examined had neutron capture element abundances reported, a large fraction of which were from their own work (see their Table 13), and so the usual risks inherent in interpreting trends in inhomogeneous datasets are not so important here.}.  While they found trends of increasing [Ba/Fe] and [Zr/Fe] with decreasing cluster age, they were not as strong as those found by D'Orazi et al.\ and Maiorca et al.\ (hereafter \citetalias{2009ApJ...693L..31D} and \citetalias{2011ApJ...736..120M}).  What is more, they found a trend of increasing [La/Fe] with {\it increasing} cluster age, in the direct opposite sense of the \citetalias{2011ApJ...736..120M} result.

Given the important role open clusters perform in tracing the chemical evolution of the disk, it is imperative to firmly establish the existence and behavior of their element abundance trends with age.  To that end, we present here an analysis of neutron capture element abundances in a sample of 19 open clusters spanning in age from $\sim$700 Myr to 10 Gyr.  Our study has many stars in common with those of Yong et al.\ (2012; hereafter \citetalias{2012AJ....144...95Y}), but we have undertaken a completely independent analysis of them.  Furthermore, only two clusters in our sample are common to the studies of \citetalias{2009ApJ...693L..31D} and \citetalias{2011ApJ...736..120M}, and therefore this study is an independent investigation of cluster abundance trends with age.  The elements investigated in this study are primarily s-process elements: Zr, Ba and La.  We also report abundances for the r-process element europium (Eu).

\begin{deluxetable*}{rrccccccccc}
\tabletypesize{\scriptsize}
\tablecolumns{11}
\tablewidth{0pt}
\tablecaption{Stellar Atmospheric Parameters \& Fe abundances\label{star_atmparam}}
\tablehead{ \colhead{Cluster} &
\colhead{Star} & \colhead{T$_{eff}$} & \colhead{log \it g} & \colhead{v$_{t}$} & \colhead{[Fe I/H]}  & \colhead{$\sigma$} & \colhead{\# lines} & \colhead{[Fe II/H]} & \colhead{$\sigma$} & \colhead{\# lines} \\
\colhead{} & \colhead{} & \colhead{(K)} & \colhead{(dex)} & \colhead{(km s$^{-1}$)} & \colhead{dex} & \colhead{dex} & 
\colhead{} & \colhead{dex} & \colhead{(dex)} & \colhead{} }
\startdata
Be 18   & 1163 & 4650 & 2.3 & 1.15 &  $-$0.30 & 0.08 & 67 & $-$0.31 & 0.12 & 8 \\
Be 18   & 1383 & 4400 & 1.9 & 1.30 & $-$0.34 & 0.09 & 71 & $-$0.33 & 0.18 & 8 \\
Be 21   & 50      & 4650 & 2.1 & 1.20 & $-$0.14 & 0.09 & 67 & $-$0.22 & 0.08 & 9 \\
Be 21   & 51      & 4520 & 1.9 & 1.40 & $-$0.28 & 0.13 & 70 & $-$0.26 & 0.11 & 9 \\
Be 22   & 414   & 4400 & 2.1 & 1.25 & $-$0.24 & 0.12 & 68 & $-$0.24 & 0.12 & 9 \\
Be 22   & 643   & 4000 & 1.1 & 1.50 & $-$0.30 & 0.33 & 64 & $-$0.26 & 0.26 & 6 \\
Be 17   & 265   & 4500 & 2.1 & 1.40 & $-$0.13 & 0.14 & 72 & $-$0.10 & 0.16 & 10 \\
Be 17   & 569   & 4300 & 1.5 & 1.50 & $-$0.11 & 0.16 & 73 & $-$0.11 & 0.16 & 11 \\
Be 17   & 1035 & 4300 & 1.7 & 1.50 & $-$0.13 & 0.15 & 73 & $-$0.19 & 0.15 & 11 \\
Be 32   & 2        & 4100 & 1.0 & 1.50 & $-$0.29 & 0.14 & 72 & $-$0.35 & 0.27 & 11 \\
Be 32   &  4       & 4100 & 1.0 & 1.50 & $-$0.33 & 0.14 & 72 & $-$0.41 & 0.26 & 11 \\
Be 32   & 16       & 4900 & 2.7 & 1.00 & $-$0.26 & 0.10 & 63 & $-$0.21 & 0.15 & 9 \\
Be 32   & 18       & 5000 & 2.7 & 1.30 & $-$0.19 & 0.13 & 66 & $-$0.21 & 0.12 & 9 \\
Be 39   & 3        & 4200 & 1.6 & 1.50 & $-$0.15 & 0.12 & 64 & $-$0.14 & 0.14 & 11 \\
Be 39   & 5        & 4450 & 1.9 & 1.40 & $-$0.11 & 0.14 & 69 & $-$0.16 & 0.18 & 11 \\
Be 39   & 12      & 4750 & 2.3 & 1.40 & $-$0.12 & 0.17 & 68 & $-$0.17 & 0.07 & 11 \\
Be 39   & 14      & 4750 & 2.3 & 1.40 & $-$0.15 & 0.15 & 70 & $-$0.13 & 0.21 & 11 \\
M 67     & 105   & 4450 & 2.1 & 1.50 & $+$0.03 & 0.12 & 64 & $-$0.01 & 0.21 & 11 \\
M 67     & 141   & 4700 & 2.4 & 1.50 & $+$0.09 & 0.12 & 65 & $+$0.07 & 0.22 & 11 \\
M 67     & 170   & 4300 & 1.7 & 1.50 & $+$0.02 & 0.14 & 60 & $+$0.00 & 0.28 & 11 \\
N 1193 & 282   & 4700 & 2.2 & 1.50 & $-$0.17 & 0.14 & 65 & $-$0.24 & 0.16 & 10 \\
N 1245 & 10     & 5000 & 2.2 & 1.20 & $-$0.01 & 0.16 & 69 & $-$0.02 & 0.21 & 11 \\
N 1245 & 125   & 4800 & 2.5 & 1.50 & $+$0.06 & 0.18 & 66 & $+$0.04 & 0.12 & 11 \\
N 1245 & 160   & 4800 & 2.5 & 1.30 & $+$0.03 & 0.17 & 63 & $+$0.03 & 0.21 & 11 \\
N 1245 & 382   & 4800 & 2.3 & 1.50 & $+$0.00 & 0.13 & 71 & $-$0.02 & 0.15 & 11 \\
N 1817 & 73     & 4850 & 2.5 & 1.40 & $-$0.04 & 0.12 & 73 & $-$0.07 & 0.13 & 11 \\
N 1817 & 79     & 5100 & 2.6 & 1.50 & $-$0.06 & 0.14 & 73 & $-$0.13 & 0.11 & 11 \\
N 188   & 532   & 4800 & 2.9 & 1.50 & $+$0.06 & 0.19 & 70 & $-$0.01 & 0.11 & 11 \\
N 188   & 747   & 4600 & 2.6 & 1.50 & $+$0.13 & 0.19 & 70 & $+$0.13 & 0.21 & 11 \\
N 188   & 919   & 4400 & 2.2 & 1.50 & $+$0.14 & 0.19 & 61 & $+$0.15 & 0.16 & 10 \\
N 188   & 1224 & 4700 & 2.9 & 1.50 & $+$0.13 & 0.18 & 71 & $+$0.12 & 0.20 & 11 \\
N 1883 &  8       & 4500 & 1.5 & 1.60 & $-$0.03 & 0.13 & 59 & $-$0.04 & 0.20 & 11 \\
N 1883 & 9        & 4600 & 1.7 & 1.60 & $-$0.04 & 0.13 & 61 & $-$0.06 & 0.19 & 11 \\
N 2141 & 1007 & 4100 & 1.2 & 1.60 & $-$0.09 & 0.16 & 59 & $-$0.07 & 0.16 & 11 \\
N 2141 & 1348 & 4100 & 1.2 & 1.50 & $-$0.09 & 0.18 & 47 & $-$0.08 & 0.18 & 8 \\
N 2158 & 4230 & 4400 & 1.5 & 1.50 & $-$0.05 & 0.15 & 58 & $-$0.11 & 0.18 & 11 \\
N 2194 & 55      & 5100 & 2.5 & 1.40 & $-$0.06 & 0.14 & 70 & $-$0.12 & 0.17 & 11 \\
N 2194 & 57      & 5100 & 2.2 & 1.40 & $-$0.06 & 0.16 & 70 & $-$0.11 & 0.18 & 11 \\
N 2355 & 144    & 5300 & 3.3 & 1.30 & $-$0.14 & 0.21 & 69 & $-$0.18 & 0.22 & 11 \\
N 2355 & 398    & 5200 & 2.1 & 1.90 & $+$0.05 & 0.13 & 71 & $+$0.01 & 0.20 & 11 \\
N 2355 & 668    & 5100 & 2.6 & 1.30 & $-$0.04 & 0.13 & 71 & $-$0.05 & 0.15 & 11 \\
N 6939 & 121    & 4200 & 1.5 & 1.60 & $+$0.00 & 0.22 & 61 & $+$0.02 & 0.18 & 11 \\
N 6939 & 190    & 5000 & 2.6 & 1.40 & $+$0.15 & 0.17 & 71 & $+$0.11 & 0.21 & 11 \\
N 6939 & 212    & 4300 & 1.7 & 1.50 & $+$0.03 & 0.15 & 65 & $+$0.07 & 0.15 & 11 \\
N 6939 & 31      & 4000 & 0.9 & 1.50 & $-$0.02 & 0.18 & 61 & $+$0.04 & 0.26 & 11 \\
N 7142 & 196    & 4500 & 2.2 & 1.40 & $+$0.08 & 0.15 & 55 & $+$0.10 & 0.14 & 11 \\
N 7142 & 229    & 4300 & 1.7 & 1.50 & $+$0.05 & 0.15 & 50 & $+$0.01 & 0.12 & 11 \\
N 7142 & 377    & 4450 & 2.0 & 1.50 & $+$0.08 & 0.13 & 56 & $+$0.03 & 0.13 & 11 \\
N 7142 & 421    & 4800 & 2.5 & 1.50 & $+$0.10 & 0.14 & 60 & $+$0.08 & 0.14 & 11 \\
PWM 4 & RGB 1 & 4000 & 1.2 & 1.50 & $-$0.18 & 0.16 & 61 & $-$0.12 & 0.30 & 8 \\
\enddata
\end{deluxetable*} 

\section{Sample}
The cluster sample used in this work is shown in Table \ref{cluster_sample}.  
Much of our stellar sample has been presented in previous work, and is comprised of echelle spectra from KPNO 4m and Hobby-Eberly telescopes (see \citealt{2005AJ....129.2725F, 2010AJ....139.1942F}, \citealt{2008AJ....135.2341J,2009AJ....137.4753J}).  This sample includes confirmed radial velocity cluster member red clump and giant stars in the clusters Be 17 (3 stars), Be 32 (2 stars), Be 39 (4 stars), M 67 (3 stars), NGC 188 (4 stars), NGC 1193 (1 star), NGC 1817 (2 stars), NGC 1883 (2 stars), NGC 2141 (2 stars\footnote{The spectrum of star 1348 was obtained by Yong et al.\ 2005 using a similar spectrograph setup to the data described here.  This spectrum was kindly given to us by D.\ Yong (2008, private communication) and was analyzed in \citet{2009AJ....137.4753J}.}), NGC 2158 (1 star) and NGC 7142 (4 stars).  The reader is referred to references above for details regarding the observations, reductions, and analysis of these data and information about the stellar targets.  

Additional high resolution spectra of stars in clusters Be 18, Be 21, Be 22, Be 32, and PWM 4 obtained with the HIRES spectrograph on Keck were kindly given to us by D.\ Yong (two stars in each cluster save one in PWM 4).  These outer disk clusters serve as an important supplement to our sample, and a comparison of the analysis here to that presented by \citetalias{2012AJ....144...95Y} provides important information about systematic differences between different cluster samples in the literature along the lines investigated by us previously (see Appendix in \citealt{2010AJ....139.1942F}, hereafter FJP10).  We refer the reader to \citetalias{2012AJ....144...95Y} for details regarding the observations and reductions of these spectra.

\begin{figure*}
\epsscale{0.8}
\plotone{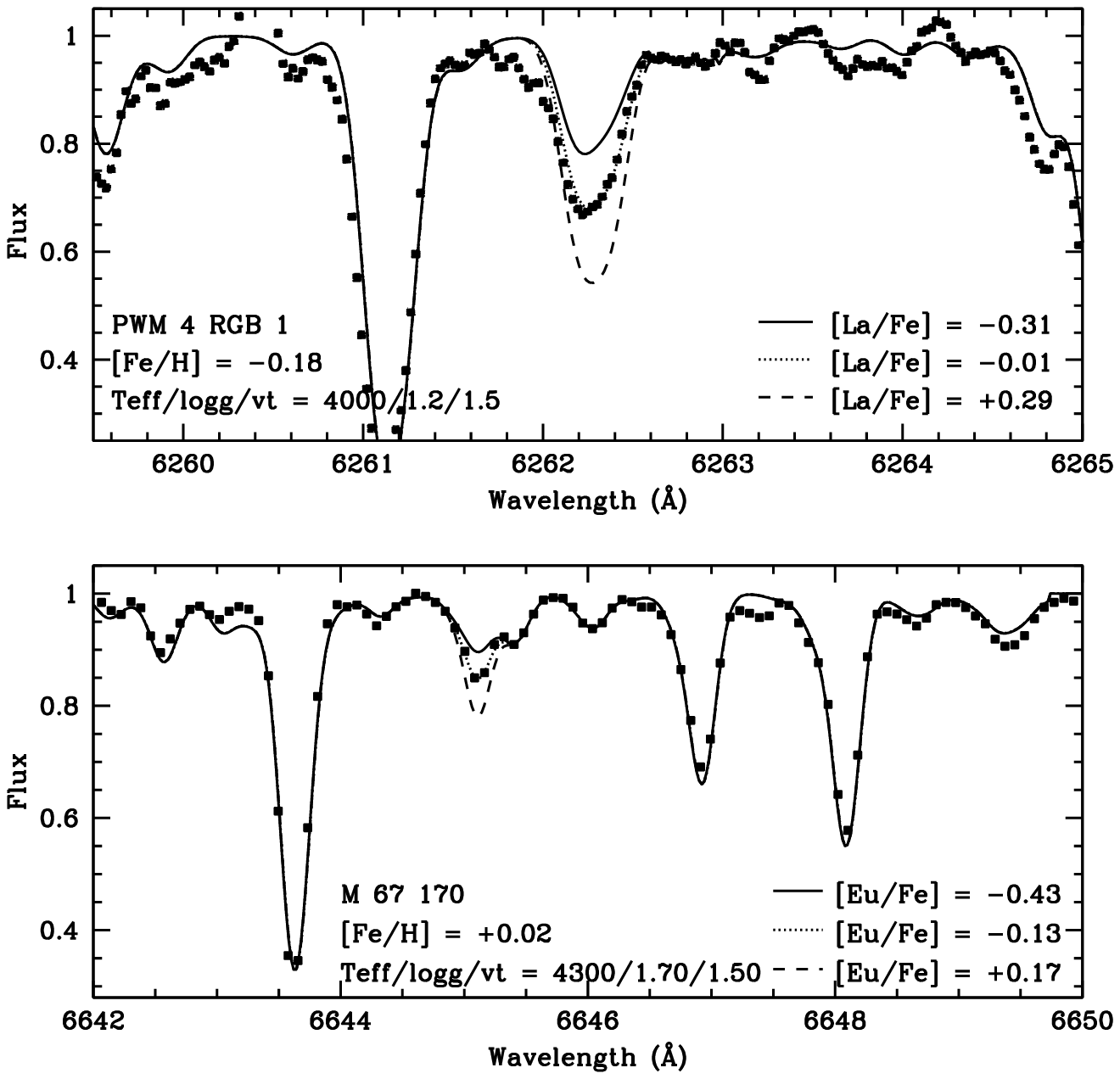}
\caption{Example syntheses of the La II 6262 feature in star PWM 4 RGB 1 (top panel) and the Eu II 6645 feature in star M 67 170 (bottom panel).}
\label{synthfig}
\end{figure*}

We obtained new observations of stars in open clusters NGC 1245, NGC 2194, NGC 2355 and NGC 6939 using the echelle spectrograph on the KPNO 4m Mayall telescope in 2010 November.  Cluster radial velocity members were selected for high resolution multi-order spectroscopic follow-up from the sample of \citet{2011AJ....142...59J}.  These data are similar to those obtained by us in previous runs (e.g., FJP10): the long red camera along with the 316 line mm$^{-1}$ grating and a 1'' slit resulted in spectra with resolving power R($\lambda$/$\Delta$$\lambda$) $\sim$ 28,000.  The full wavelength range is 5000-8100 \AA\ across 24 orders.  Typically 3-6 exposures of no more than 2700 s were obtained for each target to minimize the impact of cosmic rays.  In addition, high signal-to-noise (S/N) spectra of radial velocity standard stars and early type stars (used for subtraction of telluric lines) were obtained each night, along with the standard bias, dome flat, and arc lamp calibration frames.

These data were reduced using standard IRAF\footnote{IRAF is distributed by the National Optical Astronomy Observatory, which is operated by the Association of Universities for Research in Astronomy, Inc., under cooperative agreement with the National Science Foundation.} procedures, as described in FJP10.  After bias subtraction, scattered light removal and flat field division, individual data frames were processed through ``L. A. Cosmic"\footnote{See http://www.astro.yale.edu/dokkum/lacosmic/} to remove cosmic rays (\citealt{2001PASP..113.1420V}; spectroscopic version).  Individual orders were then extracted and converted to one-dimensional spectra and dispersion corrected.  Individual object spectra were combined to form a single, high S/N spectrum for each star and then continuum-normalized.  Signal-to-noise (S/N) ratios of the summed spectra range from 60 to 120 per pixel, with an average value of 80 per pixel.  For more information about the stellar targets, such as photometry and radial velocity, please see \citet{2011AJ....142...59J} and \citet{2007AJ....134.1216J} (for NGC 6939).  

We carried out a detailed chemical abundance analysis of these stars using the same methods and line list as described here and in our previous work (see FJP10; updated log gf's for use with MOOG2010, as described below) for elements Fe, Na, Mg, Al, Si, Ca, Ti, Cr, Co, Ni, Zr, Ba, La and Eu.  Here, we present the results for Fe and the neutron capture elements, and stellar atmospheric parameters; the other element abundances will be presented in a future paper (Jacobson et al., in prep.). 

All told, the total sample contains 50 stars in 19 clusters, with a maximum of four stars per cluster.  Figure \ref{n2141_spec} shows portions of stellar spectra with absorption features of interest identified.  Unfortunately, such small numbers of stars per cluster are insufficient to assess the true dispersions of element abundances in open clusters, and any conclusions we draw from their analysis are vulnerable to the presence of ``pathological" stars that are either non cluster members or just outliers.  We have WIYN-Hydra single-order multi-object spectroscopy of stars in nine clusters in this study, as well as five others (\citealt{2007AJ....134.1216J, 2011AJ....142...59J}, Friel et al.\ in prep.), that include La and Zr absorption features for as many as 30 confirmed members per cluster.  We will explore internal cluster abundance dispersions for La and Zr in a future paper.

\section{Analysis} 
\subsection{Atmospheric parameters and [Fe/H]}
In this work, we used the 2010 version of MOOG \citep{1973PhDT.......180S} and spherical MARCS\footnote{See http://marcs.astro.uu.se/; the models were obtained from the website 2011 July.} model atmospheres \citep{2008A&A...486..951G}.  This is a change from our previous work where we exclusively used the 2002 version of MOOG and an older grid of plane parallel MARCS models \citep{1976A&AS...23...37B}.  We note here that the change in the version of MOOG used in our analysis necessitated an update of the log gf's in our general line list (e.g., that used in FJP10), because they were determined differentially relative to Arcturus using MOOG and were therefore version-dependent.  We have modified the log gf's of all the lines in our general list to be correct for use with the 2010 version of MOOG and redetermined atmospheric parameters and element abundances for all the stars in our sample using the standard classical abundance techniques (see, e.g., FJP10).  Changes to log gf's were small with a mean difference of 0.01$\pm$0.03 (1 $\sigma$) dex and a maximum of 0.11 dex for some Ti I lines.  Relevant to this current work are the stellar atmospheric parameters, Fe and Zr abundances (see below).  Updated detailed abundances of other elements for these stars are available by request to the authors.  Differences in atmospheric parameters and element abundances were generally small -- typically within 50 K, 0.1 dex, and 0.1 \kms\ in \teff, log g and \vt, respectively (150 K, 0.3 dex and 0.2 \kms\ for the most extreme cases).  These differences are well within the estimated uncertainties of the original analysis (see also Section 3.3).  Table \ref{star_atmparam} presents the updated atmospheric parameters and iron abundances for the stars in this work.  Note that [Fe/H] values are relative to a solar log $\epsilon$(Fe) = 7.52, the default used in MOOG (see \citealt{1991AJ....102.2001S}), and that ``$\sigma$" denotes the (1 $\sigma$) standard deviations of the Fe I and Fe II abundances.

\begin{deluxetable*}{lccccccc}
\tabletypesize{\footnotesize}
\tablewidth{0pt}
\tablecaption{Line List\label{linelist}}
\tablehead{
\colhead{Element} & \colhead{Wavelength} & \colhead{E.P.} &
\colhead{log {\it gf}} & \colhead{log $\epsilon$(X)\tablenotemark{a}} & \colhead{log $\epsilon$(X)\tablenotemark{b}} &
\colhead{log $\epsilon$(X)\tablenotemark{c}} & \colhead{[X/Fe]\tablenotemark{d}}\\
\colhead{} & \colhead{(\AA)} & \colhead{} &
\colhead{} & \colhead{Sun} & \colhead{Sun} &
\colhead{Arcturus} & \colhead{Arcturus} } 
\startdata
Zr I & 6127.48 & 0.150 & $-$1.180 & 2.60 & 2.94 & 2.10 & $+$0.00\\ 
Zr I & 6134.55 & 0.000 & $-$1.426 & 2.60 & 3.02 & 2.10 & $+$0.00\\
Zr I & 6143.20 & 0.070 & $-$1.252 & 2.60 & 2.90 & 2.10 & $+$0.00\\
Ba II & 5853.70 & 0.604 & $-$1.010\tablenotemark{e} & 2.13 & 2.20 & 1.63 & $+$0.00\\
Ba II & 6141.70 & 0.704 & $-$0.070\tablenotemark{e}  & 2.13 & 2.59 & 2.30 & $+$0.67\\
Ba II & 6496.90 & 0.604 & $-$0.377\tablenotemark{e} & 2.13 &  2.41 & 2.06 & $+$0.43\\
La II & 6262.29 & 0.403 & $-$1.220\tablenotemark{e} & 1.13 & 1.22 & 0.45 & $-$0.18\\
La II & 6390.48 & 0.320 & $-$1.410\tablenotemark{e} & 1.13 & 1.20 & 0.49 & $-$0.14\\ 
Eu II & 6437.64 & 1.319 & $-$0.320\tablenotemark{e} & 0.52 & 0.53 & 0.25\tablenotemark{f} & $+$0.23\\
Eu II & 6645.10 & 1.379 & $+$0.120\tablenotemark{e} & 0.52 & 0.49 & 0.25 & $+$0.23\\
\enddata
\tablenotetext{a}{\ Literature values as described in text.}
\tablenotetext{b}{Values found via EW measures (Zr, Ba) or spectrum synthesis (La, Eu) in this work, as described in text.}
\tablenotetext{c}{This study.}
\tablenotetext{d}{Arcturus [X/Fe] ratios calculated relative to literature solar values, adopting [Fe/H] = $-$0.50.}
\tablenotetext{e}{The total log gf for the feature; log gf's for the individual components in La and Eu are presented in \citet{2001ApJ...556..452L} and \citet{2001ApJ...563.1075L}, respectively. Total log gf's and those of individual components for Ba are adopted from \citet{1998AJ....115.1640M} and \citet{2000AJ....120.2513P}.}
\tablenotetext{f}{This abundance is forced to match that for the Eu II $\lambda$6645 feature; see details in the text.}
\end{deluxetable*}

\subsection{The Neutron capture elements}
We have employed both spectrum synthesis and equivalent width (EW) measurement of absorption features for the neutron-capture elements in this study.  For Zr and Ba, we used EW measurements exclusively, while for La and Eu we used both EW measures and spectrum synthesis.  Details of the analysis for each species are given below; here, however, we briefly summarize our methods of EW and spectrum synthesis analysis.

Measurement of line EWs is straightforward and was carried out in the same way as done for Fe and the other elements in our previous work.  That is, we fit Gaussians to the line profiles interactively in IRAF (though we also measured Ba features with Voigt profiles; see below).  MOOG (version 2010) was used in our spectrum synthesis analysis, in a method identical to that employed to determine oxygen abundances in our previous work.  We carefully matched the resolution of the synthetic spectra to the observed spectra by synthesizing absorption features of other elements near the feature of interest whose abundances we determined via EW measurement (e.g., Fe, Ca, or Ni lines).  Next, we generated sets of three synthetic spectra at a time with the abundance of the element of interest varied in a stepsize 0.30 dex and the best fit abundance was determined by eye.  The uncertainty in the best fit abundance was then found by decreasing the step size between the three synthetic spectra, and then varying the level of the continuum by its approximate uncertainty, until no single synthetic spectra could be identified as the best.
 
 \begin{deluxetable}{rrccccc}
\tabletypesize{\scriptsize}
\tablecolumns{7}
\tablewidth{0pt}
\tablecaption{Individual Star [Zr/Fe] Ratios\label{star_zr}}
\tablehead{ \colhead{} & \colhead{} & \colhead{Zr $\lambda$6127} & \colhead{Zr $\lambda$6134} & \colhead{Zr $\lambda$6143} & \colhead{[Zr/Fe]} & \colhead{[Zr/Fe]}\\
\colhead{Cluster} & \colhead{Star} & \colhead{EW} & \colhead{EW} & \colhead{EW} & \colhead{mean} & \colhead{$\sigma$}} 
\startdata
Be 17   & 265  & $-$0.09 & $-$0.18 & $-$0.10 & $-$0.12 & 0.05 \\
Be 17   & 569  & $-$0.15 & $-$0.42 & $-$0.23 & $-$0.27 & 0.14 \\
Be 17   & 1035 & $+$0.20 & $-$0.14 & $-$0.02 & $+$0.01 & 0.17 \\
Be 18   & 1163 & $+$0.20 & $+$0.04 & $+$0.05 & $+$0.10 & 0.09 \\
Be 18   & 1383 & $+$0.06 & $+$0.02 & $+$0.03 & $+$0.04 & 0.02 \\
Be 21   &  50  & $+$0.24 & $+$0.17 & $+$0.17 & $+$0.19 & 0.04 \\
Be 21   &  51  & $+$0.10 & $+$0.01 & $+$0.06 & $+$0.06 & 0.05 \\
Be 22   & 414  & $+$0.27 & $+$0.06 & $+$0.10 & $+$0.14 & 0.11 \\
Be 22   & 643  & $+$0.80 & $+$0.71 & $+$0.94 & $+$0.82 & 0.12 \\
Be 32   &  2   & $-$0.08 & $+$0.05 & $-$0.01 & $-$0.01 & 0.07 \\
Be 32   &  4   & $-$0.14 & $-$0.13 & $-$0.09 & $-$0.12 & 0.03 \\
Be 32   & 16   & $+$0.07 & $+$0.21 & $+$0.18 & $+$0.15 & 0.07 \\
Be 32   & 18   & $+$0.21 & $+$0.23 & $+$0.28 & $+$0.24 & 0.04 \\
Be 39   & 3    & $-$0.05 & $-$0.15 & $+$0.08 & $-$0.04 & 0.12 \\
Be 39   & 5    & $+$0.07 & $-$0.18 & $-$0.03 & $-$0.05 & 0.13 \\
Be 39   & 12   & $+$0.05 & $+$0.09 & $-$0.05 & $+$0.03 & 0.07 \\
Be 39   & 14   & $+$0.16 & $+$0.37 & $+$0.05 & $+$0.19 & 0.16 \\
M 67     & 105 & $+$0.10 & $-$0.00 & $+$0.10 & $+$0.07 & 0.06 \\
M 67     & 141 & $-$0.03 & $-$0.06 & $-$0.02 & $-$0.04 & 0.02 \\
M 67     & 170 & $-$0.06 & $-$0.11 & $+$0.01 & $-$0.05 & 0.06 \\
N 1193 & 282   & $+$0.48 & $+$0.06 & $+$0.45 & $+$0.33 & 0.23 \\
N 1245 & 10    & $+$0.20 & $+$0.11 & $+$0.27 & $+$0.19 & 0.08 \\
N 1245 & 125   & $-$0.02 & $+$0.02 & $+$0.22 & $+$0.07 & 0.13 \\
N 1245 & 160   & $+$0.54 & $+$0.55 & $+$0.52 & $+$0.54 & 0.02 \\
N 1245 & 382   & $+$0.26 & $+$0.25 & $+$0.29 & $+$0.27 & 0.02 \\
N 1817 & 73    & $+$0.34 & $+$0.17 & $+$0.17 & $+$0.23 & 0.10 \\
N 1817 & 79    & $+$0.27 & $-$0.02 & $+$0.11 & $+$0.12 & 0.15 \\
N 188   & 532  & $+$0.28 & $+$0.06 & $+$0.45 & $+$0.26 & 0.20 \\
N 188   & 747  & $+$0.03 & $-$0.07 & $+$0.04 & $-$0.00 & 0.06 \\
N 188   & 919  & $+$0.07 & $-$0.27 & $-$0.10 & $-$0.10 & 0.17 \\
N 188   & 1224 & $+$0.13 & $+$0.13 & $+$0.19 & $+$0.15 & 0.03 \\
N 1883 & 8     & $+$0.05 & $-$0.28 & $-$0.15 & $-$0.13 & 0.17 \\
N 1883 & 9     & $-$0.01 & $-$0.11 & $-$0.13 & $-$0.08 & 0.06 \\
N 2141 & 1007  & $-$0.08 & $-$0.04 & $+$0.02 & $-$0.03 & 0.05 \\
N 2141 & 1348  & $+$0.89 & $+$0.45 & $+$0.85 & $+$0.73 & 0.24 \\
N 2158 & 4230  & $-$0.03 & $-$0.10 & $-$0.02 & $-$0.05 & 0.04 \\
N 2194 & 55    & $+$0.15 & $+$0.15 & $+$0.29 & $+$0.20 & 0.08 \\
N 2194 & 57    & $+$0.17 & $-$0.16 & $+$0.36 & $+$0.12 & 0.26 \\
N 2355 & 144   & $+$0.92 & \nodata & $+$0.59 & $+$0.76 & 0.23 \\
N 2355 & 398   & $+$0.19 & $+$0.36 & $+$0.60 & $+$0.38 & 0.21 \\
N 2355 & 668   & $+$0.23 & $+$0.47 & $+$0.28 & $+$0.33 & 0.13 \\
N 6939 & 121   & $-$0.70 & $-$0.66 & $-$0.62 & $-$0.66 & 0.04 \\
N 6939 & 190   & $+$0.55 & $+$0.48 & $+$0.82 & $+$0.62 & 0.18 \\
N 6939 & 212   & $+$0.08 & $-$0.17 & $-$0.13 & $-$0.07 & 0.13 \\
N 6939 & 31    & $+$0.28 & $+$0.11 & $+$0.21 & $+$0.20 & 0.09 \\
N 7142 & 196   & $+$0.17 & $-$0.03 & $+$0.10 & $+$0.08 & 0.10 \\
N 7142 & 229   & $+$0.29 & $+$0.04 & $+$0.20 & $+$0.18 & 0.13 \\
N 7142 & 377   & $+$0.17 & $+$0.05 & $+$0.21 & $+$0.14 & 0.08 \\
N 7142 & 421   & $+$0.14 & $-$0.10 & $+$0.24 & $+$0.09 & 0.17 \\
PWM 4 & RGB 1  & $+$0.23 & $+$0.11 & $+$0.22 & $+$0.19 & 0.07 \\
\enddata
\end{deluxetable}

\begin{deluxetable}{rrccccc}
\tabletypesize{\scriptsize}
\tablecolumns{7}
\tablewidth{0pt}
\tablecaption{Individual Star [Ba/Fe] Ratios\label{star_ba}}
\tablehead{ \colhead{} & \colhead{} & \colhead{Ba $\lambda$5853} & \colhead{Ba $\lambda$6141} & \colhead{Ba $\lambda$6496} & \colhead{[Ba/Fe]} & \colhead{[Ba/Fe]}\\
\colhead{Cluster} & \colhead{Star} & \colhead{EW} & \colhead{EW} & \colhead{EW} & \colhead{mean} & \colhead{$\sigma$}} 
\startdata
Be 17   & 265  & $+$0.04 & $-$0.24 & $+$0.05 & $+$0.04 & 0.01 \\
Be 17   & 569  & $+$0.09 & $-$0.23 & $+$0.09 & $+$0.09 & 0.00 \\
Be 17   & 1035 & $+$0.01 & $-$0.16 & $-$0.03 & $-$0.01 & 0.03 \\
Be 18   & 1163 & $+$0.42 & $+$0.11 & $+$0.36 & $+$0.39 & 0.04 \\
Be 18   & 1383 & $+$0.47 & $+$0.16 & $+$0.37 & $+$0.42 & 0.07 \\
Be 21   &  50  & $+$0.64 & $+$0.20 & $+$0.49 & $+$0.56 & 0.11 \\
Be 21   &  51  & $+$0.45 & $+$0.26 & $+$0.40 & $+$0.43 & 0.04 \\
Be 22   & 414  & $+$0.51 & $+$0.08 & $+$0.38 & $+$0.45 & 0.09 \\
Be 22   & 643  & $+$1.01 & $+$0.41 & $+$0.51 & $+$0.76 & 0.35 \\
Be 32   &  2   & $+$0.30 & $-$0.11 & $+$0.08 & $+$0.19 & 0.16 \\
Be 32   &  4   & $+$0.28 & $-$0.14 & $+$0.09 & $+$0.18 & 0.13 \\
Be 32   & 16   & $+$0.37 & $-$0.17 & $+$0.16 & $+$0.27 & 0.15 \\
Be 32   & 18   & $+$0.32 & $+$0.01 & $+$0.12 & $+$0.22 & 0.14 \\
Be 39   & 3    & $+$0.24 & $-$0.19 & $+$0.08 & $+$0.16 & 0.11 \\
Be 39   & 5    & $+$0.03 & $-$0.55 & $+$0.04 & $+$0.04 & 0.01 \\
Be 39   & 12   & $+$0.05 & $-$0.17 & $+$0.09 & $+$0.07 & 0.03 \\
Be 39   & 14   & $-$0.07 & $-$0.16 & $+$0.12 & $+$0.02 & 0.13 \\
M 67    & 105  & $+$0.07 & $-$0.31 & $+$0.00 & $+$0.04 & 0.05 \\
M 67    & 141  & $+$0.05 & $-$0.30 & $+$0.14 & $+$0.10 & 0.06 \\
M 67    & 170  & $+$0.19 & $-$0.17 & $+$0.14 & $+$0.17 & 0.04 \\
N 1193 & 282   & $+$0.15 & $-$0.05 & $+$0.13 & $+$0.14 & 0.01 \\
N 1245 & 10    & $+$0.47 & $-$0.08 & $+$0.23 & $+$0.35 & 0.17 \\
N 1245 & 125   & $+$0.70 & $+$0.15 & $+$0.31 & $+$0.51 & 0.28 \\
N 1245 & 160   & $+$0.43 & $+$0.20 & $+$0.43 & $+$0.43 & 0.00 \\
N 1245 & 382   & $+$0.44 & $+$0.05 & $+$0.34 & $+$0.39 & 0.07 \\
N 1817 & 73    & $+$0.56 & $+$0.15 & $+$0.15 & $+$0.36 & 0.29 \\
N 1817 & 79    & $+$0.32 & $+$0.04 & $+$0.04 & $+$0.18 & 0.20 \\
N 188   & 532  & $-$0.17 & $-$0.39 & $-$0.10 & $-$0.14 & 0.05 \\
N 188   & 747  & $-$0.12 & $-$0.34 & $+$0.14 & $+$0.01 & 0.18 \\
N 188   & 919  & $-$0.14 & $-$0.25 & $+$0.18 & $+$0.02 & 0.23 \\
N 188   & 1224 & $+$0.33 & $-$0.07 & $+$0.22 & $+$0.28 & 0.08 \\
N 1883 & 8     & $+$0.56 & $+$0.13 & $+$0.34 & $+$0.45 & 0.16 \\
N 1883 & 9     & $+$0.58 & $+$0.20 & $+$0.35 & $+$0.46 & 0.16 \\
N 2141 & 1007  & $+$0.37 & $-$0.01 & $+$0.17 & $+$0.27 & 0.14 \\
N 2141 & 1348  & $+$0.57 & $+$0.33 & $+$0.51 & $+$0.54 & 0.04 \\
N 2158 & 4230  & $+$0.36 & $+$0.04 & $+$0.18 & $+$0.27 & 0.13 \\
N 2194 & 55    & $+$0.68 & $+$0.13 & $+$0.44 & $+$0.56 & 0.17 \\
N 2194 & 57    & $+$0.63 & $+$0.20 & $+$0.37 & $+$0.50 & 0.18 \\
N 2355 & 144   & $+$0.47 & $+$0.17 & $+$0.46 & $+$0.46 & 0.01 \\
N 2355 & 398   & $+$0.90 & $+$0.44 & $+$0.75 & $+$0.82 & 0.11 \\
N 2355 & 668   & $+$0.52 & $+$0.17 & $+$0.40 & $+$0.46 & 0.08 \\
N 6939 & 121   & $-$0.16 & $-$0.30 & $-$0.34 & $-$0.25 & 0.13 \\
N 6939 & 190   & $+$0.98 & $+$0.46 & $+$0.60 & $+$0.79 & 0.27 \\
N 6939 & 212   & $+$0.38 & $+$0.01 & $-$0.05 & $+$0.17 & 0.30 \\
N 6939 & 31    & $+$0.53 & $+$0.07 & $+$0.04 & $+$0.29 & 0.35 \\
N 7142 & 196   & $+$0.19 & $-$0.23 & $+$0.18 & $+$0.18 & 0.01 \\
N 7142 & 229   & $+$0.24 & $-$0.09 & $+$0.21 & $+$0.22 & 0.02 \\
N 7142 & 377   & $+$0.21 & $-$0.18 & $+$0.19 & $+$0.20 & 0.01 \\
N 7142 & 421   & $+$0.20 & $-$0.22 & $+$0.16 & $+$0.18 & 0.03 \\
PWM 4 & RGB 1  & $+$0.44 & $+$0.08 & $+$0.24 & $+$0.34 & 0.14 \\
\enddata
\tablecomments{Individual line [Ba/Fe] ratios are calculated differentially relative to the solar Ba abundances listed in Table \ref{linelist}.  The mean [Ba/Fe] values are calculated using the 5853 and 6496 features only. See the text for more information.}
\end{deluxetable}

\subsubsection{Zirconium}
 The zirconium abundances are based on equivalent width (EW) measurements of three Zr I lines at $\lambda$6127, 6134, and 6143 \AA.  The log gf's of these lines were determined by measuring their EWs in the \citet{2000vnia.book.....H}\ spectrum of Arcturus and forcing the abundances to match the Zr abundance of \citeauthor{1993ApJ...404..333P} (1993; log $\epsilon$(Zr) = 2.10, normalized to the solar abundance value of \citealt{1989GeCoA..53..197A}, 2.60).  The atomic information for these lines are given in Table \ref{linelist}\footnote{Technically, five isotopes of Zr contribute to its total abundance in solar system: $^{90}$Zr, $^{91}$Zr, $^{92}$Zr, $^{94}$Zr, and $^{96}$Zr \citep{1989GeCoA..53..197A}.  Only odd isotopes are subject to hyperfine splitting, and as $^{91}$Zr comprises 11.22\%\ of the solar system abundance (and therefore only a fraction of the Zr line profiles), hyperfine and isotopic splitting are not considered here.}.
 
 Our log gf values for these Zr lines are generally smaller than those found in the literature by 0.05 to 0.15 dex (e.g., \citealt{2005AJ....130..597Y, 2012ApJ...749..175J}).  As a check on what solar abundances we would derive using these transition probabilities, we measured these lines in the solar flux atlas that accompanies the \citeauthor{2000vnia.book.....H} Arcturus atlas.  The solar abundances we derived are shown in Table \ref{linelist}, and are $\sim$0.30-0.40 dex larger than  literature values \citep[e.g.,][]{1989GeCoA..53..197A, 2009ARA&A..47..481A, 2009LanB...4B...44L}.  This indicates that our choice of log gf is possibly responsible for the systematically higher Zr abundances compared to literature values for many stars in our sample, as described below, though we note that our log gf's for Zr 6134 and 6143 agree within 0.1 dex of those of \citet{2012ApJ...749..175J}, who also found [Zr/Fe] = 0.00 for Arcturus, after adopting log $\epsilon$(Zr) = 2.60 for the Sun.  \citet{2005AJ....130..597Y}, whose log gf's for these Zr lines are 0.15 dex larger than ours, found log N(Zr) = 2.80 for the Sun, and [Zr/Fe] = $-$0.27 for Arcturus.  \citet{2000AJ....119.1239S}, whose log gf's were identical to those of \citet{2005AJ....130..597Y}, found log N(Zr) = 2.88 for the Sun, and [Zr/Fe] = -0.24 for Arcturus.  Clearly, differences in log gf's do not fully account for differences in abundances here, which, in general, should be kept in mind when comparing results of different studies in the literature.
 
\begin{deluxetable*}{rrcccccc}
\tabletypesize{\scriptsize}
\tablecolumns{8}
\tablewidth{0pt}
\tablecaption{Individual Star [La/Fe] Ratios\label{star_lafe}}
\tablehead{ \colhead{Cluster} & \colhead{Star} & \colhead{[La/Fe]$\lambda$6262} & \colhead{unc. $\lambda$6262} & \colhead{[La/Fe]$\lambda$6390} & \colhead{unc. $\lambda$6390} & \colhead{mean[La/Fe]} & \colhead{$\sigma$[La/Fe]}}
\startdata
Be 17   & 265  & $-$0.33 & 0.30 & $-$0.03 & 0.10 & $-$0.18 & 0.21 \\
Be 17   & 569  & $-$0.30 & 0.10 & $-$0.30 & 0.10 & $-$0.30 & 0.00 \\
Be 17   & 1035 & $-$0.23 & 0.07 & $-$0.08 & 0.10 & $-$0.15 & 0.11 \\
Be 18   & 1163 & $+$0.16 & 0.05 & $+$0.17 & 0.05 & $+$0.17 & 0.01 \\
Be 18   & 1383 & $+$0.03 & 0.10 & $+$0.18 & 0.06 & $+$0.10 & 0.11 \\
Be 21   &  50  & $+$0.10 & 0.05 & $+$0.13 & 0.06 & $+$0.12 & 0.02 \\
Be 21   &  51  & $+$0.14 & 0.05 & $+$0.17 & 0.05 & $+$0.16 & 0.02 \\
Be 22   & 414  & $+$0.16 & 0.05 & $+$0.20 & 0.05 & $+$0.18 & 0.03 \\
Be 22   & 643  & $+$0.07 & 0.05 & $+$0.36 & 0.07 & $+$0.21 & 0.21 \\
Be 32   &  2   & $-$0.24 & 0.05 & $-$0.12 & 0.06 & $-$0.18 & 0.08 \\
Be 32   &  4   & $-$0.33 & 0.10 & $-$0.18 & 0.07 & $-$0.26 & 0.11 \\ 
Be 32   & 16   & $+$0.10 & 0.06 & $+$0.12 & 0.07 & $+$0.11 & 0.01 \\
Be 32   & 18   & $+$0.01 & 0.06 & $+$0.03 & 0.06 & $+$0.02 & 0.01 \\
Be 39   & 3    & $-$0.23 & 0.06 & $-$0.08 & 0.06 & $-$0.15 & 0.11 \\
Be 39   & 5    & $-$0.20 & 0.10 & $+$0.02 & 0.06 & $-$0.09 & 0.16 \\
Be 39   & 12   & \nodata & \nodata & \nodata & \nodata & \nodata & \nodata \\
Be 39   & 14   & $-$0.06 & 0.10 & $+$0.04 & 0.08 & $-$0.01 & 0.07 \\
M 67     & 105 & $-$0.14 & 0.06 & $-$0.09 & 0.05 & $-$0.12 & 0.04 \\
M 67     & 141 & $-$0.15 & 0.10 & $-$0.13 & 0.07 & $-$0.14 & 0.01 \\
M 67     & 170 & $-$0.25 & 0.08 & $-$0.15 & 0.09 & $-$0.20 & 0.07 \\
N 1193 & 282   & $+$0.01 & 0.08 & $+$0.11 & 0.07 & $+$0.06 & 0.07 \\
N 1245 & 10    & $-$0.35 & 0.15 & $-$0.30 & 0.10 & $-$0.32 & 0.04 \\
N 1245 & 125   & $+$0.03 & 0.10 & $-$0.06 & 0.10 & $-$0.01 & 0.06 \\
N 1245 & 160   & $+$0.21 & 0.15 & \nodata & 0.00 & $+$0.21 & 0.15 \\ 
N 1245 & 382   & $+$0.09 & 0.10 & $+$0.09 & 0.12 & $+$0.09 & 0.00 \\
N 1817 & 73    & $+$0.10 & 0.10 & $+$0.13 & 0.08 & $+$0.12 & 0.02 \\
N 1817 & 79    & $+$0.00 & 0.10 & $+$0.15 & 0.10 & $+$0.07 & 0.11 \\
N 188   & 532  & $+$0.03 & 0.10 & $-$0.10 & 0.10 & $-$0.04 & 0.09 \\
N 188   & 747  & $-$0.19 & 0.10 & $-$0.12 & 0.10 & $-$0.15 & 0.05 \\
N 188   & 919  & $-$0.33 & 0.10 & $-$0.23 & 0.10 & $-$0.28 & 0.07 \\
N 188   & 1224 & $+$0.18 & 0.08 & $+$0.30 & 0.12 & $+$0.24 & 0.08 \\
N 1883 & 8     & $-$0.15 & 0.08 & $-$0.10 & 0.16 & $-$0.12 & 0.04 \\
N 1883 & 9     & $+$0.01 & 0.07 & $-$0.01 & 0.10 & $+$0.00 & 0.01 \\
N 2141 & 1007  & $-$0.17 & 0.05 & $-$0.06 & 0.10 & $-$0.12 & 0.08 \\
N 2141 & 1348  & $+$0.08 & 0.04 & $+$0.18 & 0.06 & $+$0.13 & 0.07 \\
N 2158 & 4230  & $-$0.16 & 0.07 & $-$0.15 & 0.08 & $-$0.15 & 0.01 \\
N 2194 & 55    & $+$0.13 & 0.10 & $+$0.06 & 0.10 & $+$0.10 & 0.05 \\
N 2194 & 57    & $+$0.00 & 0.08 & $-$0.05 & 0.10 & $-$0.03 & 0.04 \\
N 2355 & 144   & $+$0.28 & 0.10 & $+$0.38 & 0.10 & $+$0.33 & 0.07 \\
N 2355 & 398   & $+$0.07 & 0.08 & $+$0.10 & 0.08 & $+$0.09 & 0.02 \\
N 2355 & 668   & $+$0.10 & 0.10 & $+$0.13 & 0.10 & $+$0.12 & 0.02 \\
N 6939 & 121   & $-$0.41 & 0.10 & $-$0.39 & 0.10 & $-$0.40 & 0.01 \\
N 6939 & 190   & $+$0.34 & 0.10 & $+$0.49 & 0.10 & $+$0.42 & 0.11 \\
N 6939 & 212   & $-$0.19 & 0.08 & $-$0.12 & 0.09 & $-$0.15 & 0.05 \\
N 6939 & 31    & $-$0.39 & 0.05 & $-$0.29 & 0.12 & $-$0.34 & 0.07 \\
N 7142 & 196   & $-$0.24 & 0.10 & $-$0.19 & 0.10 & $-$0.21 & 0.04 \\
N 7142 & 229   & $-$0.19 & 0.07 & $-$0.06 & 0.07 & $-$0.12 & 0.09 \\
N 7142 & 377   & $-$0.21 & 0.07 & $-$0.09 & 0.06 & $-$0.15 & 0.08 \\
N 7142 & 421   & $-$0.16 & 0.06 & $-$0.16 & 0.08 & $-$0.16 & 0.00 \\
PWM 4 & RGB 1  & $-$0.01 & 0.05 & $+$0.10 & 0.06 & $+$0.05 & 0.08 \\
\enddata
\end{deluxetable*}

Note that for this present work, the EWs of $\lambda$6134 and 6143\AA\ Zr I lines have been remeasured in all stars from our previous papers and therefore may be different from values published there. The abundances from the three Zr I lines are in general good agreement with abundance differences smaller than 0.1 dex for all but the lower S/N stars (e.g., NGC 2355 stars).  For most of the stars from our previous work, remeasurement of $\lambda$6134 and 6143 features, along with the addition of $\lambda$6127 line decreased the line-by-line dispersion in abundances, and there are no systematic differences between the individual line abundances.
However, for a handful of stars, the abundance from one line differed greatly from the other two; in these cases, visual inspection of the spectra and remeasurement of the EWs did not resolve the discrepency or identify its source.  In such cases, we have included lines with discrepant abundances, even though the resulting dispersion in the mean abundance is larger.  Individual star [Zr/Fe] abundances, calculated relative to the [Fe I/H] ratios given in Table \ref{star_atmparam} are shown in Table \ref{star_zr}.  Mean cluster [Zr/Fe] abundances, with (1 $\sigma$) standard deviations, are shown in Table \ref{hires_meanabund}.

\subsubsection{Barium}
Three barium lines were measured for determination of abundances: $\lambda$5853, 6141, and 6496\AA.  All three lines are quite strong and broad, and the wings of some are blended with other spectral features (namely $\lambda$6496).
Barium features are subject to both hyperfine and isotopic broadening from seven different isotopes: $^{130}$Ba, $^{132}$Ba, $^{134}$Ba, $^{135}$Ba, $^{136}$Ba, $^{137}$Ba, and $^{138}$Ba.  According to \citet{1989GeCoA..53..197A}, these isotopes contribute 0.106\%, 0.101\%, 2.417\%, 6.592\%, 7.854\%, 11.23\%, and 71.70\% of the total solar system Ba abundance, respectively.  We have employed the line lists of \citet{1998AJ....115.1640M} (see also \citealt{2000AJ....120.2513P}), which  incorporate these isotopic fractions within the log gf's, as well as the s-process and r-process contributions to the solar system barium abundance (85\%\ and 15\%, respectively).

\begin{deluxetable*}{rrcccccc}
\tabletypesize{\scriptsize}
\tablecolumns{8}
\tablewidth{0pt}
\tablecaption{Individual Star [Eu/Fe] Ratios\label{star_eufe}}
\tablehead{ \colhead{Cluster} & \colhead{Star} & \colhead{[Eu/Fe]$\lambda$6437} & \colhead{unc. $\lambda$6437} & \colhead{[Eu/Fe]$\lambda$6645} & \colhead{unc. $\lambda$6645} & \colhead{mean [Eu/Fe]} & \colhead{$\sigma$[Eu/Fe]}}\\
\startdata
Be 17   & 265  & $+$0.02 & 0.12 & $+$0.02 & 0.08 & $+$0.02 & 0.00 \\
Be 17   & 569  & $-$0.15 & 0.15 & $-$0.15 & 0.08 & $-$0.15 & 0.00 \\
Be 17   & 1035 & $+$0.02 & 0.13 & $-$0.03 & 0.07 & $-$0.00 & 0.04 \\
Be 18   & 1163 & $+$0.11 & 0.15 & $+$0.17 & 0.07 & $+$0.14 & 0.04 \\
Be 18   & 1383 & \nodata & 0.00 & $+$0.16 & 0.07 & $+$0.16 & 0.07 \\
Be 21   &  50  & $-$0.07 & 0.15 & $-$0.12 & 0.10 & $-$0.10 & 0.04 \\
Be 21   &  51  & $+$0.14 & 0.08 & $+$0.05 & 0.08 & $+$0.10 & 0.06 \\
Be 22   & 414  & $+$0.23 & 0.13 & $+$0.18 & 0.08 & $+$0.21 & 0.04 \\
Be 22   & 643  & $+$0.41 & 0.08 & $+$0.24 & 0.08 & $+$0.32 & 0.12 \\
Be 32   &  2   & $-$0.14 & 0.10 & $-$0.02 & 0.06 & $-$0.08 & 0.08 \\
Be 32   &  4   & $+$0.00 & 0.08 & $+$0.04 & 0.06 & $+$0.02 & 0.03 \\
Be 32   & 16   & $+$0.05 & 0.10 & $+$0.08 & 0.08 & $+$0.07 & 0.02 \\
Be 32   & 18   & $+$0.13 & 0.07 & $+$0.08 & 0.05 & $+$0.11 & 0.04 \\
Be 39   & 3    & $-$0.26 & 0.12 & $-$0.09 & 0.07 & $-$0.17 & 0.12 \\
Be 39   & 5    & $+$0.00 & 0.13 & $-$0.13 & 0.07 & $-$0.07 & 0.09 \\
Be 39   & 12   & $+$0.01 & 0.10 & $-$0.09 & 0.15 & $-$0.04 & 0.07 \\
Be 39   & 14   & $-$0.01 & 0.13 & $-$0.01 & 0.08 & $-$0.01 & 0.00 \\
M 67     & 105 & $-$0.14 & 0.10 & $-$0.07 & 0.07 & $-$0.11 & 0.05 \\
M 67     & 141 & $-$0.10 & 0.10 & $-$0.05 & 0.07 & $-$0.08 & 0.04 \\
M 67     & 170 & $-$0.28 & 0.10 & $-$0.13 & 0.07 & $-$0.21 & 0.11 \\
N 1193 & 282   & $+$0.06 & 0.10 & $-$0.04 & 0.10 & $+$0.01 & 0.07 \\
N 1245 & 10    & \nodata  & \nodata & \nodata & \nodata & \nodata & \nodata \\
N 1245 & 125   & \nodata  & \nodata & \nodata & \nodata & \nodata & \nodata \\
N 1245 & 160   & \nodata  & \nodata & \nodata & \nodata & \nodata & \nodata \\
N 1245 & 382   & \nodata  & \nodata & \nodata & \nodata & \nodata & \nodata \\
N 1817 & 73    & $-$0.12 & 0.15 & $-$0.07 & 0.10 & $-$0.10 & 0.04 \\
N 1817 & 79    & $-$0.20 & 0.15 & $-$0.10 & 0.10 & $-$0.15 & 0.07 \\
N 188   & 532  & $+$0.08 & 0.15 & $-$0.22 & 0.10 & $-$0.07 & 0.21 \\ 
N 188   & 747  & $-$0.44 & 0.20 & $-$0.04 & 0.12 & $-$0.24 & 0.28 \\ 
N 188   & 919  & $-$0.53 & 0.20 & $-$0.13 & 0.10 & $-$0.33 & 0.28 \\ 
N 188   & 1224 & $-$0.34 & 0.20 & $+$0.16 & 0.10 & $-$0.09 & 0.35 \\ 
N 1883 & 8     & $-$0.30 & 0.13 & $-$0.30 & 0.15 & $-$0.30 & 0.00 \\
N 1883 & 9     & $-$0.09 & 0.10 & $-$0.20 & 0.10 & $-$0.15 & 0.08 \\
N 2141 & 1007  & $-$0.22 & 0.10 & $-$0.14 & 0.08 & $-$0.18 & 0.06 \\
N 2141 & 1348  & $-$0.22 & 0.10 & $-$0.12 & 0.10 & $-$0.17 & 0.07 \\
N 2158 & 4230  & $-$0.16 & 0.10 & $-$0.11 & 0.10 & $-$0.14 & 0.04 \\
N 2194 & 55    & $-$0.30 & 0.18 & $-$0.10 & 0.10 & $-$0.20 & 0.14 \\
N 2194 & 57    & $-$0.30 & 0.15 & $-$0.30 & 0.15 & $-$0.30 & 0.00 \\
N 2355 & 144   & $+$0.28 & 0.15 & \nodata & 0.00 & $+$0.28 & 0.15 \\ 
N 2355 & 398   & $-$0.01 & 0.10 & $-$0.16 & 0.10 & $-$0.09 & 0.11 \\
N 2355 & 668   & $-$0.35 & 0.30 & $-$0.17 & 0.10 & $-$0.26 & 0.13 \\
N 6939 & 121   & $-$0.31 & 0.15 & $-$0.21 & 0.10 & $-$0.26 & 0.07 \\
N 6939 & 190   & \nodata & \nodata & $-$0.13 & 0.10 & $-$0.13 & 0.10 \\
N 6939 & 212   & $-$0.34 & 0.17 & $-$0.11 & 0.08 & $-$0.23 & 0.16 \\
N 6939 & 31    & $-$0.26 & 0.15 & $-$0.34 & 0.10 & $-$0.30 & 0.06 \\
N 7142 & 196   & $-$0.29 & 0.10 & $-$0.19 & 0.10 & $-$0.24 & 0.07 \\
N 7142 & 229   & $-$0.46 & 0.15 & $-$0.21 & 0.10 & $-$0.34 & 0.18 \\
N 7142 & 377   & $-$0.54 & 0.30 & $-$0.19 & 0.10 & $-$0.36 & 0.25 \\
N 7142 & 421   & $-$0.31 & 0.15 & $-$0.21 & 0.10 & $-$0.26 & 0.07 \\
PWM 4 & RGB 1  & $+$0.07 & 0.10 & $+$0.07 & 0.08 & $+$0.07 & 0.00 \\
\enddata
\end{deluxetable*}

Abundances from these lines were calculated using measured EWs.  To determine the best method to measure EWs, we measured these lines in the \citet{2000vnia.book.....H} Arcturus and solar flux atlases by fitting both Gaussian and Voigt profiles to them.  The Ba abundances corresponding to the Gaussian line profiles are indicated in Table \ref{linelist}.  Voigt profiles resulted in abundances 0.2-0.6 dex larger than that those for the Gaussian fits, and much more deviant from the solar value found in the literature (e.g., 2.13 -- \citealt{1989GeCoA..53..197A}, 2.24 -- \citealt{1996ASPC...99..117G}).  We also measured Ba abundances in the Sun from these lines via spectrum synthesis and find values  more consistent with the solar value: log $\epsilon$(Ba) = 2.13, 2.23, and 2.26 for $\lambda$5853, 6141, and 6496, respectively.  Using the same three Ba lines as this study, \citet{2000AJ....119.1239S} found [Ba/Fe] = $-$0.05 for Arcturus; \citet{2005AJ....130..597Y} found [Ba/Fe] = $+$0.09 based on the $\lambda$5853 line only.  Our value for $\lambda$5853 is consistent with these values.

As a result of this analysis of the Sun and Arcturus, we measured all three Ba features in our sample stars by fitting Gaussian profiles, as these values are more consistent with those both from the spectrum synthesis analysis and from the literature.  Given the especially large discrepancy between the spectrum synthesis and EW abundance for the $\lambda$6141 line in the Sun and Arcturus, and that its abundance is also often discrepant from those of the other Ba lines in our program stars, we have opted not to include this line in the calculation of mean Ba abundances for our stellar sample, though we include it in our results for completeness.  Table \ref{star_ba} shows the [Ba/Fe] ratios for each star calculated from the individual lines, with the mean and standard deviations calculated from the $\lambda$5853 and 6496 lines only.  {\it Note that these [Ba/Fe] ratios are calculated relative to our determined solar Ba abundances in column 6 of Table \ref{linelist}, to minimize systematics due to the zero-point of our abundance scale.}   Mean cluster [Ba/Fe] abundances, with (1 $\sigma$) standard deviations, are shown in Table \ref{hires_meanabund}.

 \begin{deluxetable*}{lccccccccccc}
\tabletypesize{\footnotesize}
\tablewidth{0pt}
\tablecaption{Cluster mean abundances\label{hires_meanabund}}
\tablehead{
\colhead{Cluster} & \colhead{[Fe/H]} & \colhead{$\sigma$} & \colhead{[Zr/Fe]} & \colhead{$\sigma$} & 
\colhead{[Ba/Fe]} & \colhead{$\sigma$} & \colhead{[La/Fe]} & \colhead{$\sigma$} & \colhead{[Eu/Fe]} &
\colhead{$\sigma$} & \colhead{\# Stars}
}
\startdata
Be 17          & $-$0.12 & 0.01 & $-$0.13 & 0.14 & $+$0.04 & 0.05 & $-$0.21 & 0.08 & $-$0.04 & 0.09 & 3 \\
Be 18          & $-$0.32 & 0.03 & $+$0.07 & 0.04 & $+$0.41 & 0.02 & $+$0.14 & 0.05 & $+$0.15 & 0.01 & 2 \\
Be 21          & $-$0.21 & 0.10 & $+$0.13 & 0.09 & $+$0.50 & 0.09 & $+$0.14 & 0.03 & $+$0.00 & 0.14 & 2 \\
Be 22\tablenotemark{1}          & $-$0.24 & 0.12 & $+$0.14 & 0.11 & $+$0.45 & 0.09 & $+$0.18 & 0.03 & $+$0.21 & 0.04 & 1 \\
Be 32          & $-$0.27 & 0.06 & $+$0.07 & 0.16 & $+$0.22 & 0.04 & $-$0.08 & 0.17 & $+$0.03 & 0.08 & 4 \\
Be 39          & $-$0.13 & 0.02 & $+$0.03 & 0.11 & $+$0.07 & 0.06 & $-$0.08 & 0.07 & $-$0.07 & 0.07 & 4 \\
M 67           & $+$0.05 & 0.04 & $-$0.01 & 0.07 & $+$0.10 & 0.07 & $-$0.15 & 0.04 & $-$0.13 & 0.07 & 3 \\
NGC 188   & $+$0.12 & 0.04 & $+$0.08 & 0.16 & $+$0.04 & 0.17 & $-$0.06 & 0.22 & $-$0.18 & 0.12 & 4 \\
NGC 1193 & $-$0.17 & 0.13 & $+$0.33 & 0.23 & $+$0.14 & 0.01 & $+$0.06 & 0.07 & $+$0.01 & 0.07 & 1 \\
NGC 1245 & $+$0.02 & 0.03 & $+$0.27 & 0.20 & $+$0.42 & 0.07 & $-$0.01 & 0.23 & \nodata & \nodata & 4 \\
NGC 1817 & $-$0.05 & 0.01 & $+$0.18 & 0.08 & $+$0.27 & 0.13 & $+$0.10 & 0.04 & $-$0.13 & 0.04 & 2 \\
NGC 1883 & $-$0.04 & 0.02 & $-$0.11 & 0.04 & $+$0.46 & 0.02 & $-$0.06 & 0.08 & $-$0.23 & 0.11 & 2 \\
NGC 2141 & $-$0.09 & 0.01 & $+$0.50 & 0.71 & $+$0.41 & 0.19 & $+$0.01 & 0.18 & $-$0.18 & 0.01 & 2 \\
NGC 2158 & $-$0.05 & 0.14 & $-$0.05 & 0.04 & $+$0.27 & 0.13 & $-$0.15 & 0.01 & $-$0.14 & 0.04 & 1 \\
NGC 2194 & $-$0.06 & 0.01 & $+$0.16 & 0.06 & $+$0.41 & 0.19 & $+$0.04 & 0.09 & $-$0.25 & 0.07 & 2 \\
NGC 2355 & $-$0.04 & 0.10 & $+$0.49 & 0.24 & $+$0.58 & 0.21 & $+$0.18 & 0.13 & $-$0.02 & 0.28 & 3 \\
NGC 6939\tablenotemark{2} & $+$0.00 &  0.03 & $-$0.31 & 0.31 & $+$0.07 & 0.28 & $-$0.30 & 0.13 & $-$0.26 & 0.04 & 3 \\
NGC 7142 & $+$0.08 & 0.02 & $+$0.12 & 0.05 & $+$0.20 & 0.02 & $-$0.16 & 	0.04 & $-$0.30 & 0.06 & 4 \\
PWM 4       & $-$0.18 &  0.16 & $+$0.19 & 0.07 & $+$0.34 & 0.14 & $+$0.05 & 0.08 & $+$0.07 & 0.01 & 1 \\
\enddata
\tablenotetext{1}{[X/Fe] ratios calculated without star 643.}
\tablenotetext{2}{[X/Fe] ratios calculated without star 190.}
\end{deluxetable*}

\subsubsection{Lanthanum}
$^{139}$La comprises 99.911\%\ of the solar system La abundance \citep{1989GeCoA..53..197A}, and so La absorption features are subject to hyperfine broadening.  We have adopted the line list of \citet{2001ApJ...556..452L}, which includes data for individual hyperfine lines, for the analysis of the $\lambda$6262 and $\lambda$6390 \AA\ features.  We have determined La abundances for all stars in this study via both spectrum synthesis and EW measurement of the lines.  We take the spectrum synthesis abundance values to be more reliable, since fitting the line features with Gaussians to determine an equivalent width may not necessarily well fit the broad La features in some stars.  Figure \ref{synthfig} shows an example spectrum synthesis of the La II 6262 feature in one of our program stars.

First though, it is useful to confirm the zero-point of our La abundances.  To do so, we have determined abundances of the Sun and Arcturus via spectrum synthesis, again using the \citet{2000vnia.book.....H} Arcturus atlas and the accompanying solar flux atlas.  Using the \citet{2001ApJ...556..452L} line list, we find log $\epsilon$(La) = 1.22 and 1.20 for $\lambda$6262 and $\lambda$6390 \AA\ for the Sun, respectively (see Table \ref{linelist}).  These values, while slightly larger than the log $\epsilon$(La) = 1.13 found by \citet{2001ApJ...556..452L},  are in good agreement with solar values found by  \citet{1989GeCoA..53..197A} and meteoritic values found by \citet{2009LanB...4B...44L}\footnote{As pointed out by the referee, \citet{2001ApJ...556..452L} used the  empirical model atmosphere of the Sun from \citet{1974SoPh...39...19H} in their analysis.  This may at least partly explain the differences between our determined La abundances and their values.}.  The individual line La abundances for Arcturus also agree very well (within 0.04 dex of each other); however, our mean [La/Fe] ratio for Arcturus, $-$0.16$\pm$0.03, is 0.10 dex lower than that found by \citet{2012ApJ...749..175J} using the same line list and spectrum.  Our value is even lower than that found by \citet{2005AJ....130..597Y} and \citet{2000AJ....119.1239S} for Arcturus: [La/Fe] = $+$0.03$\pm$0.09 and [La/Fe] = $+$0.21, respectively (though we note that this difference in [La/Fe] values is largely driven by a $\sim$0.2 dex difference in [Fe/H] found for Arcturus between \citet{2000AJ....119.1239S} and other studies cited here).  
We adopt the \citet{2001ApJ...556..452L} solar La abundance of 1.13 to calculate [La/Fe] ratios for this study.  Individual star [La/Fe] ratios as determined via spectrum synthesis are shown in Table \ref{star_lafe}.  The abundance uncertainty for each line given in this Table represents the measurement uncertainty in the synthesis technique, as described in the introduction of this Section.  Mean cluster [La/Fe] abundances, with (1 $\sigma$) standard deviations, are shown in Table \ref{hires_meanabund}.  (For cluster abundances based on the measures of a single star only, the 1 $\sigma$ value represents the uncertainty (standard deviation) in its abundance from Table \ref{star_lafe}.)

Abundances from each La II feature were also determined using measured EWs and the {\it blends} driver in MOOG, to correctly account for the hyperfine structure of the lines.  Differences between synthesis and EW abundances for each line were plotted against [Fe/H], \teff, and log g to look for systematic trends.  For the 6262 feature, the EW abundances were $\sim$0.07 smaller than the synthesis abundances, with a standard deviation of $\sim$0.15 dex.  While no trend with [Fe/H] is present, the difference between EW and synthesis abundances grows larger (that is, EW abundances get increasingly smaller relative to the synthesis values) with increasing \teff\ and log g.  For the 6390 feature, the difference between EW and synthesis abundances (in the sense EW $-$ synthesis) is $\sim$ $+$0.13, with a standard deviation of $\sim$0.12 dex.  No trend is present versus [Fe/H], \teff, or log g.

\begin{deluxetable*}{llcccc}
\tabletypesize{\footnotesize}
\tablewidth{0pt}
\tablecaption{Uncertainties in abundance ratios due to atmospheric parameters\label{unc}}
\tablehead{
\colhead{} & \colhead{} & \colhead{T$+$100 K} & \colhead{log g$+$0.2 dex} & \colhead{[M/H]$+$0.2 dex} & \colhead{v$_{t}$$+$0.2 km s$^{-1}$}\\
\colhead{Star} & \colhead{El.} & \colhead{$\Delta$abund.} & \colhead{$\Delta$abund.} & \colhead{$\Delta$abund.} & \colhead{$\Delta$abund.}  } 
\startdata
N6939 31 & [Fe/H] & $-$0.03 & $+$0.06 & $+$0.05 & $-$0.13\\
                   & [Zr/Fe] & $+$0.19 & $-$0.01 & $+$0.01 & $-$0.09\\
                   & [Ba/Fe]$_{5853}$ & $+$0.03 & $+$0.01 & $+$0.06 & $-$0.09\\
                   & [Ba/Fe]$_{6496}$ & $+$0.04 & $+$0.01 & $+$0.06 & $-$0.03\\ 
                   & [La/Fe]$_{6262}$ & $+$0.05 & $+$0.06 & $-$0.06 & $+$0.13\\
                   & [La/Fe]$_{6390}$ & $+$0.00 & $+$0.10 & $-$0.17 & $+$0.11\\
                   & [Eu/Fe]$_{6437}$ & $-$0.02 & $+$0.05 & $-$0.09 & $+$0.01\\
                   & [Eu/Fe]$_{6645}$ & $+$0.02 & $+$0.03 & $+$0.02 & $+$0.12\\
\hline                   
N1817 79 & [Fe/H] & $+$0.09 & $+$0.01 & $+$0.01 & $-$0.07\\
                   & [Zr/Fe] & $+$0.06 & $-$0.01 & $-$0.01 & $+$0.07\\
                   & [Ba/Fe]$_{5853}$ & $-$0.07 & $+$0.06 & $+$0.04 & $-$0.15\\
                   & [Ba/Fe]$_{6496}$ & $-$0.06 & $+$0.05 & $+$0.06 & $-$0.13\\
                   & [La/Fe]$_{6262}$ & $-$0.06 & $+$0.08 & $-$0.01 & $+$0.06\\
                   & [La/Fe]$_{6390}$ & $-$0.03 & $+$0.09 & $-$0.04 & $+$0.07\\
                   & [Eu/Fe]$_{6437}$ & $-$0.08 & $+$0.10 & $-$0.25 & $+$0.14\\
                   & [Eu/Fe]$_{6645}$ & $-$0.10 & $+$0.08 & $+$0.05 & $+$0.06\\
\hline                   
N1883 9 & [Fe/H] & $+$0.06 & $+$0.03 & $+$0.02 & $-$0.10\\
                 & [Zr/Fe] & $+$0.14 & $-$0.02 & $-$0.03 & $+$0.08\\
                 & [Ba/Fe]$_{5853}$ & $-$0.06 & $+$0.05 & $+$0.06 & $-$0.14\\
                 & [Ba/Fe]$_{6496}$ & $-$0.04 & $+$0.04 & $+$0.06 & $-$0.07\\
                 & [La/Fe]$_{6262}$ & $-$0.04 & $+$0.07 & $+$0.00 & $+$0.09\\
                 & [La/Fe]$_{6390}$ & $+$0.00 & $+$0.11 & $-$0.11 & $+$0.09\\
                 & [Eu/Fe]$_{6437}$ & $-$0.08 & $+$0.08 & $-$0.11 & $+$0.09\\
                 & [Eu/Fe]$_{6645}$ & $-$0.07 & $+$0.07 & $+$0.05 & $+$0.09\\
\hline
N7142 196 & [Fe/H] & $+$0.03 & $+$0.04 & $+$0.04 & $-$0.10\\
                     & [Zr/Fe] & $+$0.18 & $-$0.03 & $-$0.05 & $+$0.04\\
                     & [Ba/Fe]$_{5853}$ & $-$0.01 & $+$0.03 & $+$0.03 & $-$0.15\\
                     & [Ba/Fe]$_{6496}$ & $-$0.01 & $+$0.01 & $+$0.06 & $-$0.06\\
                     & [La/Fe]$_{6262}$ & $-$0.01 & $+$0.07 & $-$0.07 & $+$0.10\\
                     & [La/Fe]$_{6390}$ & $-$0.02 & $+$0.09 & $-$0.10 & $+$0.09\\
                     & [Eu/Fe]$_{6437}$ & $-$0.06 & $+$0.08 & $-$0.12 & $+$0.06\\
                     & [Eu/Fe]$_{6645}$ & $-$0.05 & $+$0.04 & $+$0.02 & $+$0.09\\
\hline
PWM 4 1 & [Fe/H] & $-$0.03 & $+$0.05 & $+$0.03 & $-$0.13\\
                 & [Zr/Fe] & $+$0.20 & $+$0.00 & $-$0.01 & $-$0.06\\
                 & [Ba/Fe]$_{5853}$ & $+$0.04 & $+$0.02 & $+$0.04 & $-$0.07\\
                 & [Ba/Fe]$_{6496}$ & $+$0.04 & $+$0.01 & $+$0.04 & $-$0.02\\
                 & [La/Fe]$_{6262}$ & $+$0.05 & $+$0.05 & $+$0.00 & $+$0.13\\
                 & [La/Fe]$_{6390}$ & $+$0.02 & $+$0.07 & $-$0.08 & $+$0.10\\
                 & [Eu/Fe]$_{6437}$ & $-$0.01 & $+$0.05 & $-$0.03 & $-$0.01\\
                 & [Eu/Fe]$_{6645}$ & $+$0.01 & $+$0.04 & $+$0.02 & $+$0.11\\
\hline
Be 22 414 & [Fe/H] & $+$0.03 & $+$0.04 & $+$0.03 & $-$0.10\\
                    & [Zr/Fe] & $+$0.16 & $-$0.03 & $-$0.05 & $+$0.03\\
                    & [Ba/Fe]$_{5853}$ & $-$0.01 & $+$0.02 & $+$0.02 & $-$0.15\\
                    & [Ba/Fe]$_{6496}$ & $+$0.00 & $+$0.00 & $+$0.04 & $-$0.05\\
                    & [La/Fe]$_{6262}$ & $-$0.01 & $+$0.05 & $-$0.02 & $+$0.10\\
                    & [La/Fe]$_{6390}$ & $-$0.01 & $+$0.07 & $-$0.10 & $+$0.09\\
                    & [Eu/Fe]$_{6437}$ & $-$0.05 & $+$0.06 & $-$0.04 & $+$0.05\\
                    & [Eu/Fe]$_{6645}$ & $-$0.04 & $+$0.05 & $+$0.02 & $+$0.09\\

\enddata
\end{deluxetable*}

\subsubsection{Europium}
Like lanthanum, europium is subject to strong hyperfine splitting.  However, while La features are predominantly composed of a single isotope, Eu features are also subject to splitting between two isotopes.  We have adopted the line list for the Eu II $\lambda$6437 and $\lambda$6645 \AA\ features from \citet{2001ApJ...563.1075L}, assuming the standard solar system isotopic mix of 47.8\%\ $^{151}$Eu and 52.2\%\ $^{153}$Eu \citep{1989GeCoA..53..197A}.  

It is important to state that, according to \citet{2001ApJ...563.1075L}, the $\lambda$6437 \AA\ feature is significantly blended with a silicon feature at $\lambda$6437.71 \AA.  We have attempted to reduce the impact of this blend and recover reliable Eu abundances from this feature in the following way.  We measured the Eu abundance for Arcturus from the Eu II $\lambda$6645 line and set that as the ``true" Eu abundance: log $\epsilon$(Eu) = 0.25.  We then generated a synthetic spectrum of the $\lambda$6437 feature using Arcturus's atmospheric parameters, the Eu abundance from the $\lambda$6645 line, and the Si abundance we have adopted for Arcturus (and which served as the basis for our astrophysical log gf's; we also used an E.P. of 5.863 eV, retrieved from VALD\footnote{See http://vald.astro.univie.ac.at/$\sim$vald/php/vald.php, \citet{2000BaltA...9..590K}.}).  We then altered the log gf of the $\lambda$6437.71 Si feature until the synthetic spectrum matched the Arcturus spectrum, and used this log gf ($-$2.30) for the Si feature in the synthesis line list for Eu $\lambda$6437.  Lastly, in the spectrum synthesis analysis of our sample stars, we set the Si abundance of this blend feature to match that for each star as determined from the EW analysis of other Si lines in its spectrum.

 \begin{deluxetable*}{llccccl}
\tabletypesize{\footnotesize}
\tablewidth{0pt}
\tablecaption{Cluster abundances from the literature\label{cluster_lit}}
\tablehead{
\colhead{Cluster} & \colhead{Ref} & \colhead{[Zr/Fe]} & \colhead{[Ba/Fe]} & \colhead{[La/Fe]} & \colhead{[Eu/Fe]} & \colhead{Notes}
 }
\startdata
Be 18 & Yong et al.\ (2012) &$-$0.21$\pm$0.13 & $+$0.30$\pm$0.06 & $+$0.34$\pm$0.08 & $+$0.30$\pm$0.06 & 2 stars \\
Be 21 & Yong et al.\ (2012) & $-$0.11$\pm$0.08 & $+$0.59$\pm$0.01 & $+$0.56$\pm$0.01 & $+$0.31$\pm$0.08 & 2 stars \\
Be 22 & Yong et al.\ (2012) & $-$0.14 & $+$0.61$\pm$0.05 & $+$0.38$\pm$0.02 & $+$0.26$\pm$0.10 & 2 stars \\
Be 32 & Yong et al.\ (2012) & $-$0.06$\pm$0.07 & $+$0.29$\pm$0.12 & $+$0.44$\pm$0.02 & $+$0.31$\pm$0.11 & 2 stars \\
Be 32 & D'Orazi et al.\ (2009) & \nodata & $+$0.24$\pm$0.15 & \nodata & \nodata & 9 stars\\
Be 32 & Carrera \& Pancino (2011) & \nodata & $+$0.51$\pm$0.12 & $-$0.14$\pm$0.07 & \nodata & 2 stars \\
M 67 & Yong et al.\ (2005) & $-$0.28$\pm$0.03 & $-$0.02$\pm$0.04 & $+$0.11$\pm$0.02 & $+$0.06$\pm$0.03 & 3 stars \\
M 67 & Tautvai\v{s}ien\.{e} et al.\ (2000)& $-$0.17$\pm$0.09 & $+$0.07$\pm$0.09 & $+$0.13$\pm$0.09 & $+$0.07$\pm$0.09 & 9 stars \\
M 67 & D'Orazi et al.\ (2009) & \nodata & $+$0.04$\pm$0.05 & \nodata & \nodata & 10 stars \\
M 67 & Maiorca et al.\ (2011) & $+$0.04$\pm$0.05 & \nodata & $+$0.06$\pm$0.04 & \nodata & 10 stars \\
M 67 & Pancino et al.\ (2010) & \nodata & $+$0.25$\pm$0.06 & $+$0.05$\pm$0.06 & \nodata & 3 stars\\
NGC 1817 & Reddy et al.\ (2012) & $+$0.08$\pm$0.05 & $+$0.13 & $+$0.12$\pm$0.03 & $+$0.13 & 3 stars\\
NGC 2141 & Yong et al.\ (2005) & $+$0.63$\pm$0.06 & $+$0.91 & $+$0.57$\pm$0.07 & $+$0.17 & 1 star \\
PWM 4 & Yong et al.\ (2012) &  $+$0.11$\pm$0.06 & $+$0.46 & $+$0.22$\pm$0.07 & $+$0.12 & 1 star \\
\enddata
\end{deluxetable*}

As for La, we verified our method of analysis with the Sun.  Spectrum synthesis using the solar flux atlas from \citet{2000vnia.book.....H} resulted in log $\epsilon$(Eu) = 0.53 and 0.49 for $\lambda$6437 and $\lambda$6645, respectively.  These values are consistent to those found by \citet{2001ApJ...563.1075L} for the same lines: 0.55 and 0.54.  For Arcturus, our [Eu/Fe] = $+$0.23 is in good agreement with that found by \citet{2012ApJ...749..175J} and \citet{2005AJ....130..597Y}, again using similar analysis techniques ([Eu/Fe] = $+$0.29 in both studies, from only the $\lambda$6645 line).  Our value of [Eu/H] = $-$0.27 for Arcturus is also in good agreement with \citet{2000AJ....119.1239S}, who found [Eu/H] = $-$0.30.  Table \ref{star_eufe} presents the [Eu/Fe] ratios for our sample as determined from spectrum synthesis.  Again, the abundance uncertainties associated with each line represents the measurement uncertainty from the synthesis technique.  We note that the S/N ratios of the spectra of all four NGC 1245 stars were too low to determine reliable abundances for the relatively weak ($<$20 m\AA\ in these stars) Eu II features.  Mean cluster [Eu/Fe] abundances, with (1 $\sigma$) standard deviations, are shown in Table \ref{hires_meanabund}.  (As for La, the 1 $\sigma$ value for clusters based on the analysis of a single star represents the uncertainty (standard deviation) in its abundance from Table \ref{star_eufe}.)  The bottom panel of Figure \ref{synthfig} shows an example synthesis of the Eu II 6645 feature.

It can be seen in Table \ref{star_eufe} that the Eu abundances determined from the $\lambda$6437 feature generally agree well (i.e, with differences comparable to or smaller than the measurement uncertainties) with that of the $\lambda$6645 feature for most stars.  However, larger discrepancies are evident for some stars, especially those in clusters NGC 188 and NGC 7142.  While differing line abundances could be due to uncertainties in continuum placement or less-than-optimal S/N ratios for some stars, the stars exhibiting the largest abundance discrepancies are generally the more metal rich stars in the sample, with [Fe/H] values of $\sim$0.10 or higher.  For such metal-rich stars, our fix for neutralizing the Si blend in the $\lambda$6437 feature may break down.

Eu abundances were determined using measured EWs and the {\it blends} driver in MOOG, similar to lanthanum.  As before, differences between EW and synthesis abundances were plotted against stellar [Fe/H], \teff\ and log g.  Both Eu features show a trend of increasing difference with increasing [Fe/H], but no trend with \teff\ or log g (maybe a slight trend with log g for $\lambda$6437).  The mean difference for both lines is $\sim$0.1 dex, with the smallest abundance difference for metal poor stars ($\sim$0 dex) and the largest abundance difference ($\sim$0.2 dex) for the more metal-rich.

\subsection{Error analysis}
As in our previous work, we estimate the uncertainties in atmospheric parameters to be 100 K in \teff, 0.2 dex in log g, and 0.2 \kms\ in microturbulent velocity.  We have calculated uncertainties in [Fe/H] and [X/Fe] ratios due to these uncertainties in atmospheric parameters for six stars in our sample that span a wide range in atmospheric parameters and metallicities.  The results are presented in Table \ref{unc}.  We did not consider the covariance terms in the error analysis, rather we treated the uncertainty an each parameter as though they were independent of one another.  The uncertainty calculations for La and Eu lines were performed using EWs, not spectrum synthesis.
We include here also the abundance changes due to an uncertainty of 0.2 dex in the model atmosphere [M/H] value.  This is meant to be indicative only of the impact that such an uncertainty would have; given that we always matched the [M/H] of the model to the iron abundance of each star, this uncertainty is overly conservative.  As can be seen, [X/Fe] uncertainties are generally smaller than 0.10 dex, but occasionally as large as 0.20 dex.

\begin{figure*}
\epsscale{0.8}
\plotone{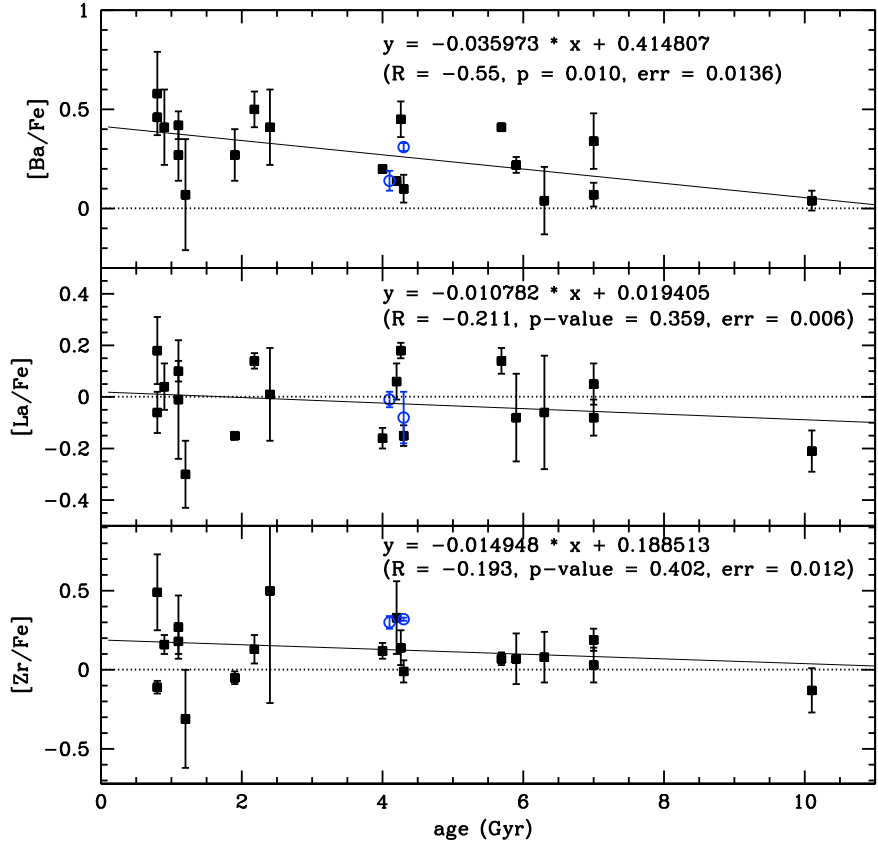}
\caption{[Ba/Fe] (top panel), [La/Fe] (middle panel), and [Zr/Fe] (bottom panel) ratios for our cluster sample (filled squares) as a function of cluster age. The open blue circles represent clusters Be 20 and Be 29 from Yong et al.\ (2005) shifted on to our abundance scale (see text for details).  The equations for least squares fits to the data are indicated in all panels, along with the R-values, p-values, and errors of the fit.}
\label{sfe_age}
\end{figure*}

\begin{figure*}
\epsscale{0.7}
\plotone{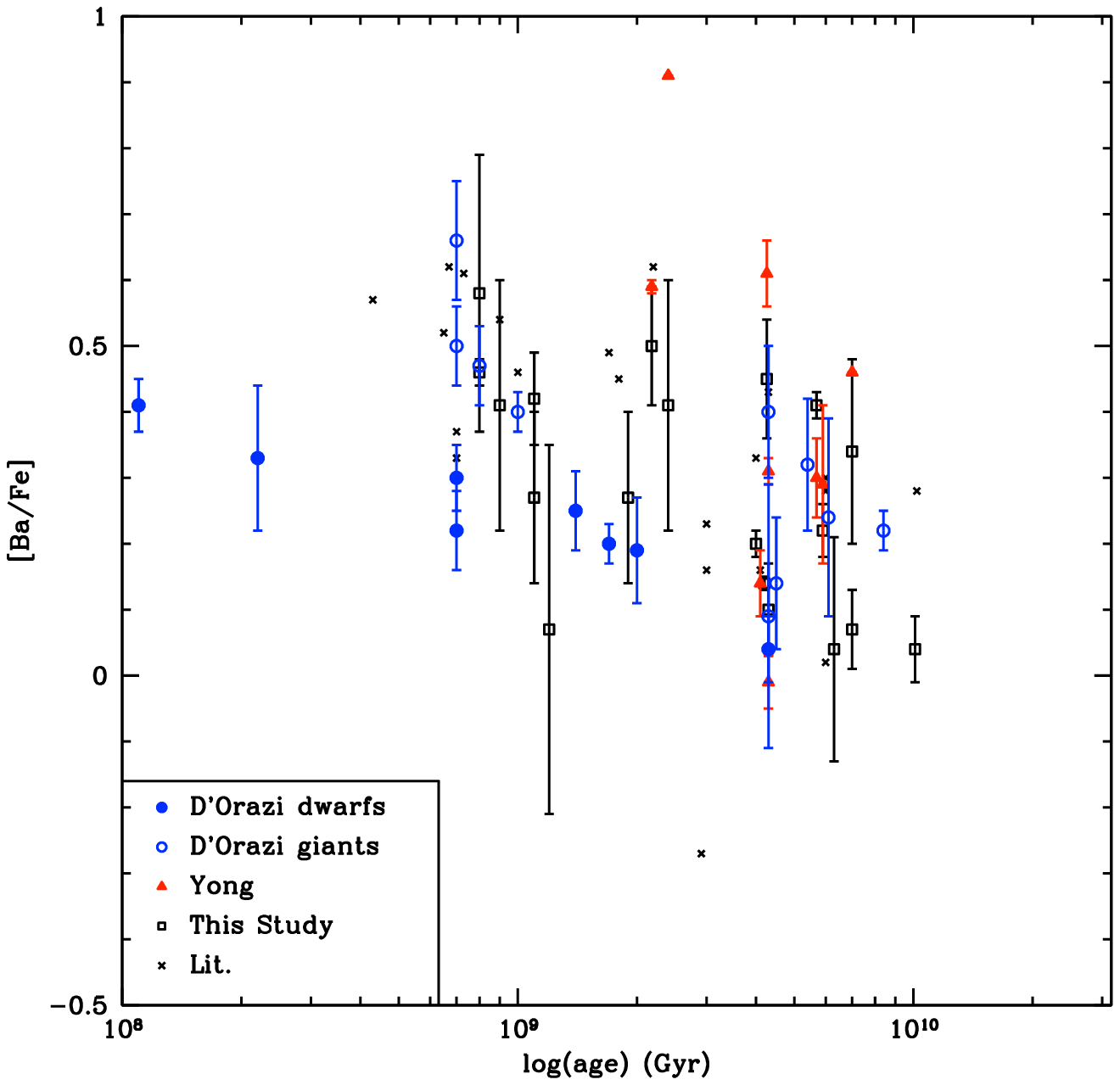}
\caption{[Ba/Fe] versus log(age) for our clusters (open squares), those of Yong et al.\ (2005, 2012; red triangles), and the sample of \citetalias{2009ApJ...693L..31D} (dwarfs -- filled blue circles, giants -- open blue circles).  Other clusters from the literature are represented by small crosses.  See text for more information.}
\label{ba_age}
\end{figure*}

To estimate the uncertainty in the Eu abundance calculated from the $\lambda$6437 line due to the Si blend, we generated a synthetic spectrum for each of the stars in Table \ref{unc} with the Si abundance varied by the line-by-line dispersion in its Si abundance as measured from EWs of Si lines.  The best fit Eu abundance was then compared to the abundances given in Table \ref{star_eufe}.  Line-by-line dispersions in Si abundances ranged from 0.08 to 0.16 dex.  Corresponding changes in the Eu abundance ranged from 0 to $\sim$0.15 dex, with a median change of only 0.02 dex.

\subsection{Anomalous stars}
A few clusters contain stars that exhibit strikingly different n-capture abundance patterns.  

Be 22: Star 643 is much more enhanced in [Zr/Fe] and [Ba/Fe] than star 414.  In fact, the abundances of several other elements in 643 differ significantly from those of 414, and the general line-by-line dispersion in element abundances is also higher (even for iron; we also note that \citetalias{2012AJ....144...95Y} found similar behavior).  Given that 643 is high up on the red giant branch relative to 414 (e.g., Figures 5 \& 6 in \citetalias{2012AJ....144...95Y}), we have taken the abundances of star 414 to be more reliable, and therefore we have adopted the abundances of star 414 as representative of those for Be 22 as a whole.  The abundances reported for Be 22 in Table \ref{hires_meanabund} are exactly those of star 414.

NGC 2141: Star 1348 shows systematically larger enhancements in [Zr/Fe], [Ba/Fe] and [La/Fe] than star 1007, though their [Eu/Fe] ratios are nicely consistent (see Figure \ref{n2141_spec}). They have very similar  atmospheric parameters, metallicity, and other element abundances.  We also note that \citet{2005AJ....130..597Y} also found high [Zr/Fe], [Ba/Fe], and [La/Fe] ratios for star 1348.  Given the similar evolutionary states of these two stars, we do not choose one and exclude the other as we have done for Be 22.  Instead, we have averaged the abundances of the two stars together, and therefore the mean abundances of this cluster show large dispersions in the subsequent discussion.

\begin{figure*}
\epsscale{0.8}
\plotone{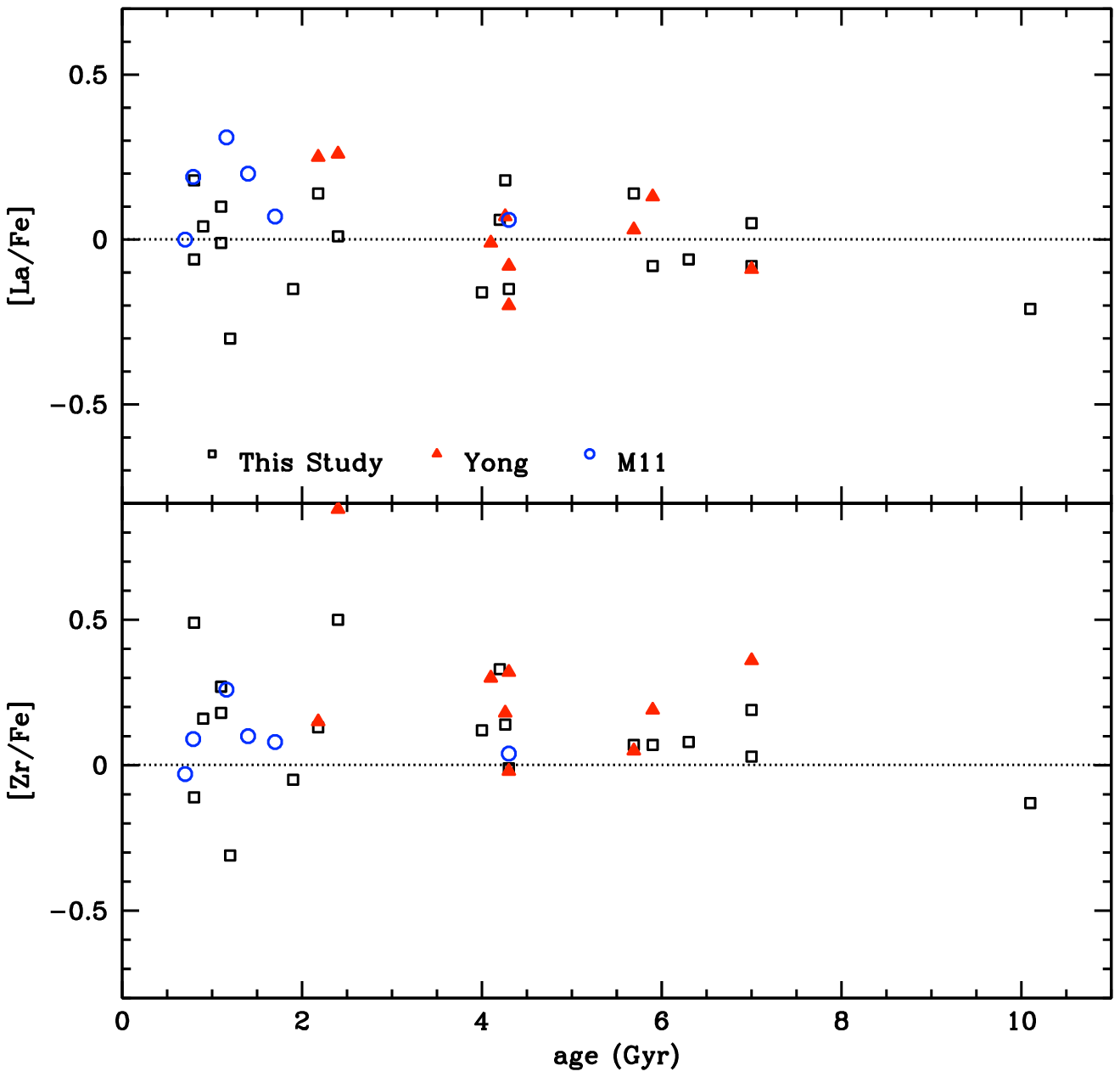}
\caption{[La/Fe] (top) and [Zr/Fe] (bottom) abundance ratios versus age for our sample (open squares), those of Yong et al.\ (2005, 2012; red triangles) shifted to our abundance scale, and those of \citetalias{2011ApJ...736..120M} (blue open circles).  Error bars have been omitted for clarity.  As described in the text, the [s/Fe] for clusters from \citetalias{2011ApJ...736..120M} fit nicely within the distribution of our sample.}
\label{sfe_m_age}
\end{figure*}

NGC 6939: Stars in this cluster show a pattern in which one star, 190, is consistently enhanced in [X/Fe] ratios,  and another (121) is consistently under-abundant relative to the remaining two stars.  Star 190 resides in the clump and shows [s/Fe] ratios at least 0.4 dex larger than the other stars.  As a result we do not include it in cluster mean abundance calculations.  Star 121 is more of a puzzle, as it is not systematically different from the other two cluster members for the other elements.  However, we note that the spectroscopic \teff\ that we found by minimizing Fe I abundances versus line E.P. is 300 K cooler than the photometric \teff\ (Jacobson et al., in prep.), while the photometric and spectroscopic temperatures for the other stars agree within 50 K.  Therefore, it is possible that star 121's discrepent [X/Fe] ratios are due to offsets in its atmospheric parameters.

Other stars show less pathological differences from their counterparts.  For example, star 398 in NGC 2355 is 0.3 dex more enhanced in [Ba/Fe], while star 1224 in NGC 188 also shows larger ($\sim$0.2-0.3 dex) enhancements in [Ba/Fe] and [La/Fe] than the other NGC 188 stars, even after taking into account line-to-line abundance dispersions.  The four stars in NGC 1245 show large variations in [La/Fe], but this may be the result of relatively low S/N ratios for the spectra (indeed, they were deemed too poor to reliably measure Eu abundances).  In general, star-to-star variations in abundances of the r-process element Eu are much smaller than for the s-process elements.  For example, the pairs of stars in Be 22 and NGC 2141 that differ in [s/Fe] ratios have similar [Eu/Fe] ratios, and the Eu abundance for star 190 in NGC 6939 is consistent within the uncertainties of the other stars in the cluster.  Instead, the two stars in Be 21  differ in [Eu/Fe] by 0.2 dex, but at a 2-$\sigma$ level.  Not surprisingly, the star-to-star scatter is larger for NGC 188 and NGC 2355 stars, with standard deviations in [Eu/Fe] of 0.12 and 0.28 dex, respectively.

\begin{figure*}
\epsscale{0.8}
\plotone{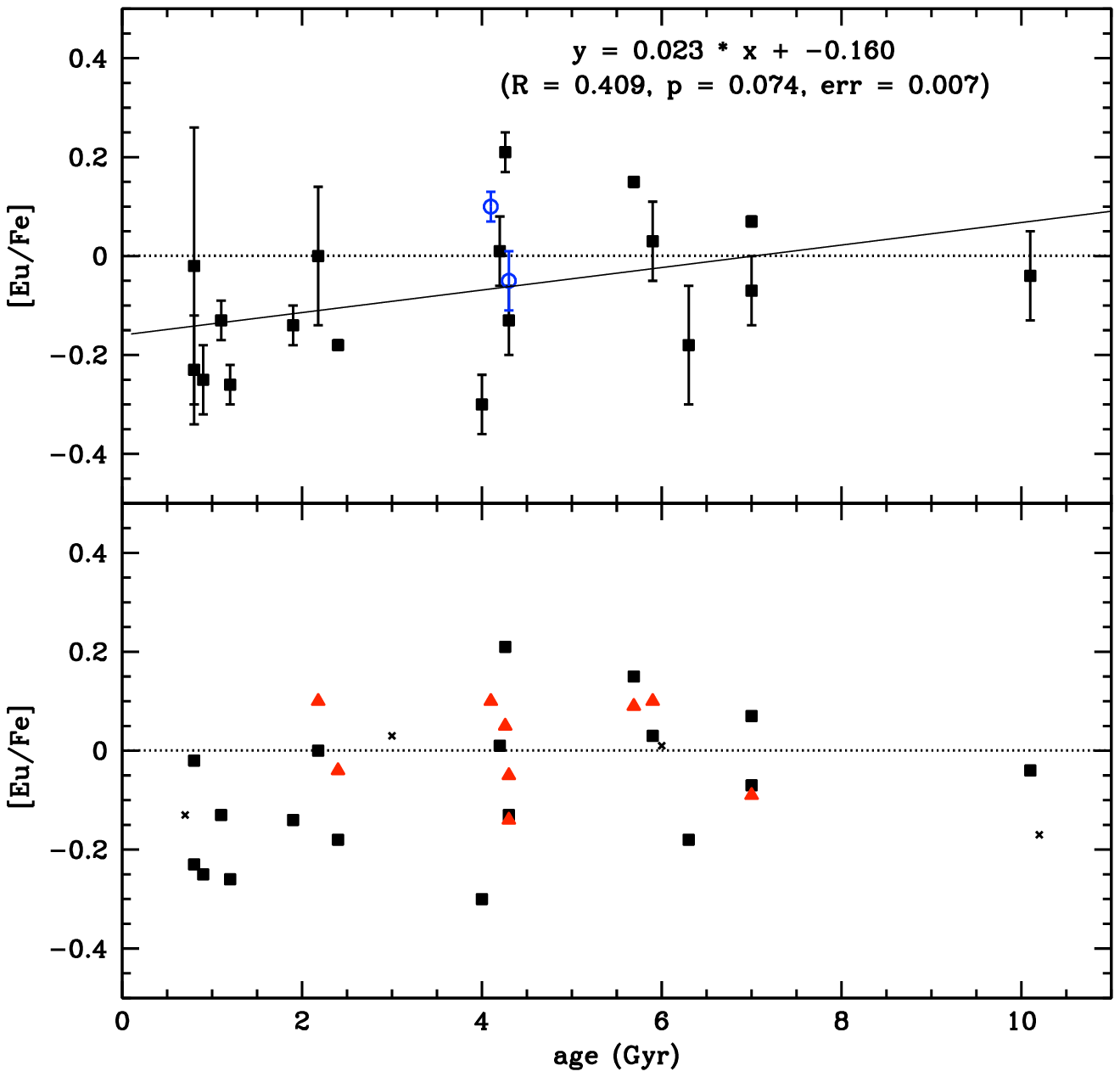}
\caption{Top panel: [Eu/Fe] versus age; symbols same as in Figure \ref{sfe_age}.  Bottom panel: [Eu/Fe] for our cluster sample (filled squares), along with those of Yong et al.\ (2005, 2012; red triangles) shifted to our abundance scale, and other clusters from the literature (crosses).}
\label{eu_age}
\end{figure*}

\subsection{Comparison to other studies}
To our knowledge, abundances of neutron capture elements are available in the literature for eight clusters in our study.  Table \ref{cluster_lit} gives a summary of these literature results.  Not surprisingly, M 67 is the most well-studied with two sets of measures for Zr, Ba, La, and Eu \citep{2005AJ....130..597Y, 2000A&A...360..499T}, and three additional studies of Ba and La abundances (\citetalias{2009ApJ...693L..31D}; \citetalias{2011ApJ...736..120M}; \citealt{2010A&A...511A..56P}).  For this cluster, our [Ba/Fe] ratio is consistent with the roughly scaled-solar values found by these studies, save for the [Ba/Fe] = $+$0.25$\pm$0.06 value found by \citet{2010A&A...511A..56P}.  Our [La/Fe] ratio of $-0.15 \pm 0.04$ for M 67 is systematically lower than those of other studies, which found [La/Fe] $\sim$0.05 to 0.13 dex.  Our [Eu/Fe] measures for M 67 stars 
are also lower than the findings of other studies in the literature (e.g., $+$0.07$\pm$0.09 -- \citealt{2000A&A...360..499T}, $+$0.06$\pm$0.03 -- \citealt{2005AJ....130..597Y}).  As for Zr, our value is consistent with that of \citetalias{2011ApJ...736..120M}, but is larger than those values found by Yong et al.\ and Tautvai\v{s}ien\.{e} et al.

The next best-studied cluster is Be 32, with literature abundances again from \citetalias{2012AJ....144...95Y}, \citetalias{2009ApJ...693L..31D}, \citetalias{2011ApJ...736..120M}, and \citet{2011A&A...535A..30C}.  As for M 67, our mean cluster [Ba/Fe] ratio agrees well with those of \citetalias{2012AJ....144...95Y} and \citetalias{2009ApJ...693L..31D} (within 0.05 dex), but is $\sim$0.25 dex lower than that of Carrera \& Pancino.    Only Carrera \& Pancino and \citetalias{2012AJ....144...95Y} presented [La/Fe] ratios for Be 32; our value is in good agreement with the former (within 0.06 dex), but is lower than the latter by 0.5 dex!  Similarly our [Eu/Fe] ratio is 0.3 dex lower than that of \citetalias{2012AJ....144...95Y}, but our scaled-solar [Zr/Fe] is consistent with their findings.

 The recent work by \citet{2012MNRAS.419.1350R} also presented neutron capture abundances for NGC 1817.  Here we find consistent [La/Fe] ratios (within 0.02 dex) for the cluster, while our [Ba/Fe] and [Zr/Fe] ratios are consistent within the uncertainties.  Our Eu abundances, however, are very different: their [Eu/Fe] = $+$0.13 is 0.26 dex larger than our value.

In addition to the clusters mentioned above, clusters NGC 2141, Be 18, Be 21, Be 22 and PWM 4 are in common with Yong et al.\ (2005, 2012) -- recall the stars and spectra for the three Berkeley clusters and star 1348 in NGC 2141 are those of Yong et al., and we have in addition two M 67 stars in common (105 and 141).  A star by star comparison of the 12 stars common between our two samples shows that differences between our abundances (in the sense This Study $-$ Yong) for [Ba/Fe], [La/Fe], [Eu/Fe] and [Zr/Fe] are $+$0.04$\pm$0.19, $-$0.29$\pm$0.12, $-$0.16$\pm$0.13, and $+$0.26$\pm$0.10 dex, respectively.  The standard deviations of this difference for La, Eu, and Zr are comparable to individual star uncertainties, while the scatter for Ba is larger, indicating that the differences between our studies are systematic for La, Eu and Zr, but not for Ba.  For the discussions that follow, we have made use of these systematic differences to place two additional clusters studied by Yong et al.\ (2005), Be 20 and Be 29, on to our cluster abundance scale.  

\section{Discussion}
In order to minimize possible systematic uncertainties in the cluster abundance results as much as possible, we limit our discussion to the results of this work with the addition of Be 20 and Be 29 from \citet{2005AJ....130..597Y}, as just discussed.  For the first discussion, that of abundance trends with age and R$_{gc}$, our sample size is comparable to those of \citetalias{2009ApJ...693L..31D} and \citetalias{2011ApJ...736..120M} and yet is almost completely distinct from them, with only two clusters in common to each (M 67 and Be 32).  Therefore, this dataset provides an excellent opportunity to look for the anticorrelation of cluster s-process abundance ([s/Fe]) with cluster age found by \citetalias{2009ApJ...693L..31D} and \citetalias{2011ApJ...736..120M}, and further explored in, \citet{2012ApJ...747...53M}.

\begin{figure*}
\epsscale{0.8}
\plotone{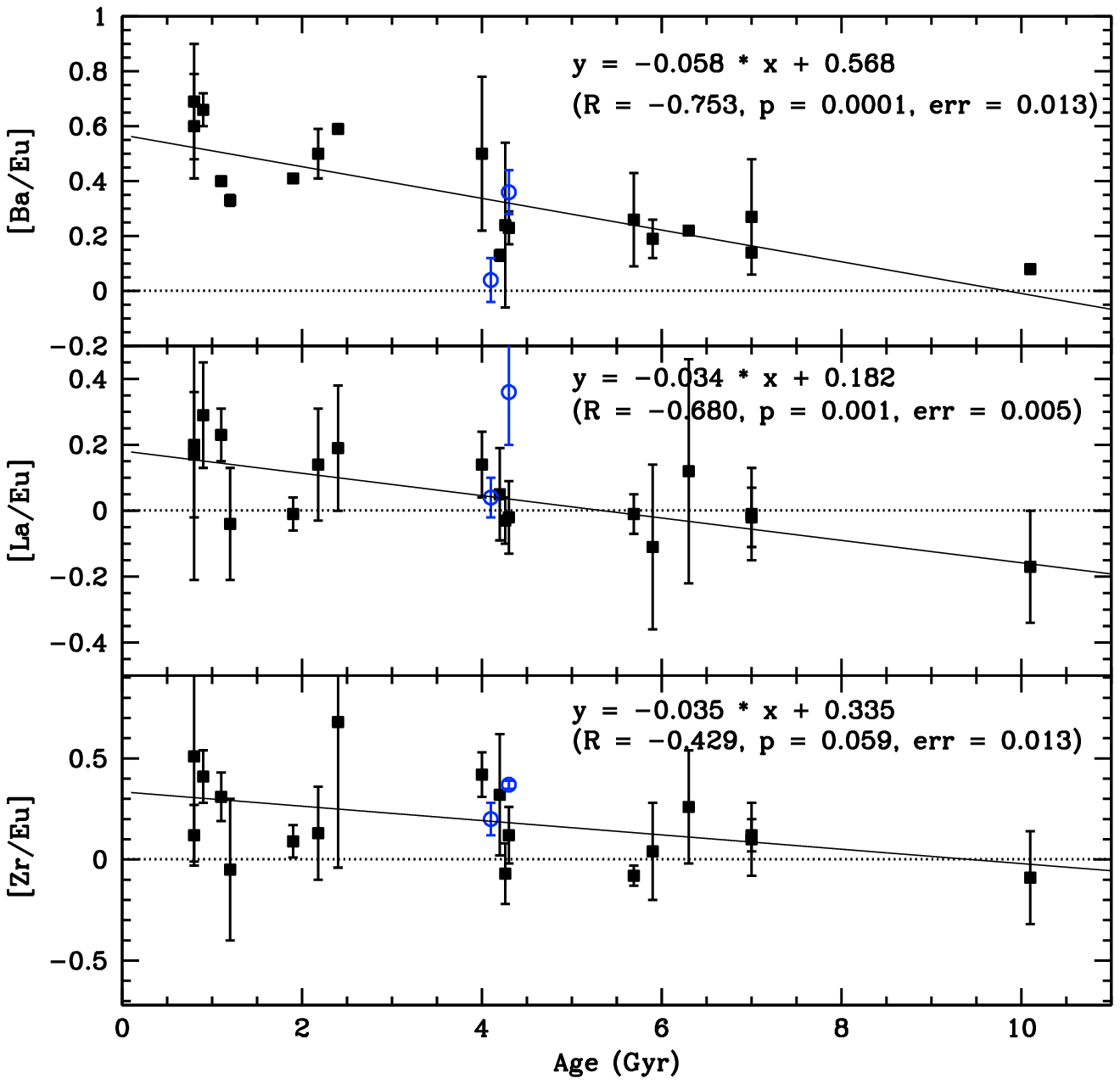}
\caption{[Ba/Eu] (top panel), [La/Eu] (middle panel), and [Zr/Eu] (bottom panel) versus age.  Symbols are the same as in Figure \ref{sfe_age}.}
\label{seu_age}
\end{figure*}

\subsection{Cluster n-capture abundance trends with age}
The first s-process element shown to have enhancements anti-correlated with open cluster age was barium (\citetalias{2009ApJ...693L..31D}), so we begin with it here.  The top panel of Figure \ref{sfe_age} shows mean cluster [Ba/Fe] ratios from Table \ref{hires_meanabund} plotted against cluster age (taken from \citealt{2004A&A...414..163S}).  In this figure, our sample is represented with squares, while open blue circles represent the outer disk clusters Be 20 and Be 29 from Yong et al.\ (2005), shifted to our abundance scale.  A simple linear regression analysis on the distribution of the points resulted in the equations for lines of best fit through the data also printed in each panel.  Also included are the R-value, standard error of the estimate for regression (err; i.e., the degree of scatter of the points around the regression line), and the two-tailed probability estimate (p-value), a measure of the statistical significance of the slope.  Conventionally, p-values smaller than 0.05 or 0.01 are taken to be rejections of the null hypothesis, which in this case is that there is no trend in [X/Fe] ratio with cluster age.  As can be seen in Figure \ref{sfe_age}, our data do show a statistically significant trend of  increasing [Ba/Fe] with decreasing cluster age, as shown in the sample of \citetalias{2009ApJ...693L..31D}.  If the oldest cluster, Be 17, is excluded, the p-value increases from p=0.010 to p = 0.048, decreasing the significance of the trend.

Results from our sample also agree with those of \citetalias{2012AJ....144...95Y}, who found a similar slope of [Ba/Fe] with age, of $-0.03 \pm 0.01$, from their sample of 10 clusters combined with another 22 measurements taken from the literature.   As they note, this trend is not as pronounced as that found in D'Orazi et al.  The advantage of the work presented here is that it includes 3 more clusters with ages greater than 6 Gyr.  Notably, our [Ba/Fe] ratios for these old clusters are all roughly solar, reinforcing the significance of the trend of decreasing Ba with cluster age.  Nevertheless, the correlation shows substantial scatter about the mean trend.  This can also be seen in Figure \ref{ba_age}, which replicates Figure 1 in \citetalias{2009ApJ...693L..31D} (open and filled blue circles), with our sample (open squares) also included.  Here, we include the full cluster sample of Yong et al.\ (2005, 2012; red triangles), along with clusters from the literature as compiled in \citetalias{2012AJ....144...95Y} (their Table 13) represented by small crosses.  We follow \citetalias{2012AJ....144...95Y} here in that we treat each measure of an individual cluster independently, and therefore some clusters are represented by multiple points in this figure.  
It can be seen that although each sample spans a slightly different range in age, they follow the same general trend of increasing [Ba/Fe] with decreasing age.  This is even seen in the inhomogeneous cluster sample drawn from the literature.

\begin{figure*}
\epsscale{0.8}
\plotone{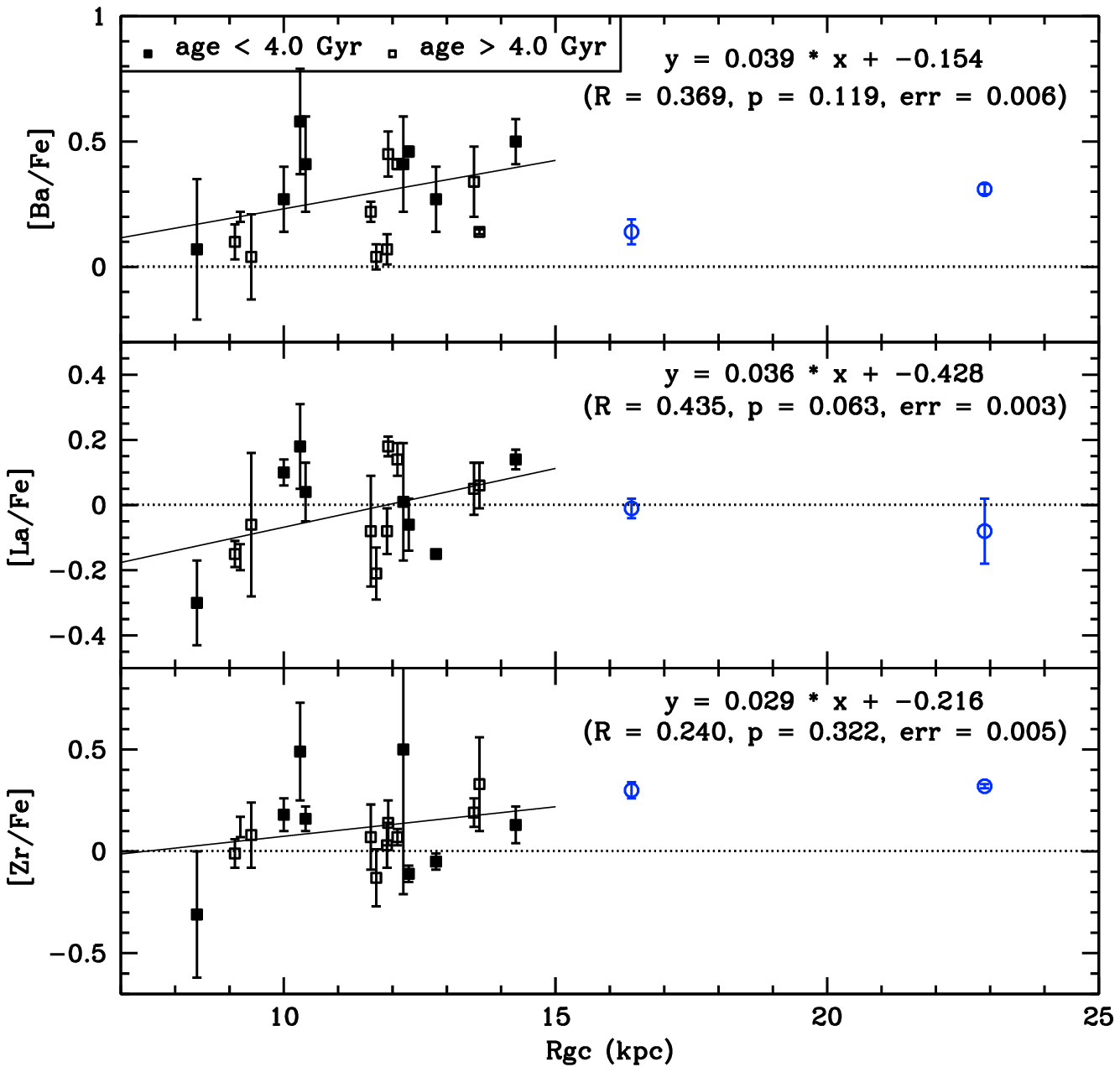}
\caption{[s/Fe] ratios as a function Galactocentric distance.  Symbols are the same as in Figure \ref{sfe_age}.  Clusters are distinguished by age: clusters younger than 4 Gyr (filled symbols), clusters 4 Gyr and older (open symbols).  In each panel, the line of best fit for {\it all} clusters within \rgc $<$ 15 kpc is indicated, along with its equation, R-value, p-value, and error.}
\label{sfe_rgc}
\end{figure*}

The middle panel of Figure \ref{sfe_age} shows cluster [La/Fe] versus age for our sample, again with two outer disk clusters from Yong et al.\ (2005) adjusted to our abundance scale.  \citetalias{2011ApJ...736..120M} found a trend of increasing [La/Fe] with decreasing cluster age, but this was based on only six clusters in their sample for which they could determine La abundances (the trend of [s/Fe] with cluster age is much more convincing for the s-process elements Ce and Y, for which they have abundances for all 19 clusters in their sample).  We note that the clusters for which they determined La abundances covered a much more limited age range (and a primarily younger one) than that presented here.  The larger sample of  Figure \ref{sfe_age} clearly shows no trend of [La/Fe] with cluster age: the p-value is quite large.  Indeed, while Be 17 shows a subsolar [La/Fe] ratio that drives some slope to the data, the rest of the clusters show an essentially flat distribution, again with a scatter of roughly 0.3 dex.  

\citetalias{2012AJ....144...95Y}, on the contrary, found a positive correlation of [La/Fe] with age, with a slope of 0.05 $\pm$ 0.01, based on their determinations for 10 clusters and literature values for 6 more.  However, recall that our values of [La/Fe] for stars in common to the \citetalias{2012AJ....144...95Y} analysis showed a systematic offset of 0.29 dex, ours being lower.  When the \citetalias{2012AJ....144...95Y} values are adjusted to our abundance scale, the trend with age disappears entirely, yielding results consistent with ours, as can be seen in the top panel of Figure \ref{sfe_m_age}.  Here, the sample of \citetalias{2012AJ....144...95Y}, shifted to our abundance scale, is indicated by red triangles, while the six clusters in \citetalias{2011ApJ...736..120M} are blue circles  
We conclude that none of the available data from these larger samples (that is, ours and \citetalias{2012AJ....144...95Y}) show evidence for age trends in La abundance, once systematic differences are taken into account, and the distribution with age of the \citetalias{2011ApJ...736..120M} sample is consistent with this observation.

\citetalias{2011ApJ...736..120M} also claimed a trend in their cluster [Zr/Fe] ratios versus cluster age, but as for La this was based on a  subset of their sample of only 6 primarily younger clusters.  We see no trend in our cluster [Zr/Fe] ratios versus age; indeed, the distribution is very similar to that of [La/Fe] (bottom panel of Figure \ref{sfe_age}).  Interestingly, the dispersion of [Zr/Fe] ratios with age is fairly tight at 0.1-0.2 dex for clusters around 6-7 Gyr of age, with the dispersion increasing with decreasing cluster age.  We note the cluster with [Ba/Fe] $\sim$0.70 with the huge errorbars is NGC 2141, whose large and discrepant Zr abundances have already been discussed.  

\citetalias{2012AJ....144...95Y} also found a slight anticorrelation of [Zr/Fe] with cluster age, with a slope of $-$0.03 $\pm$ 0.01, based again on their sample plus results from the literature, the majority of which were taken from our previous papers.   We  note that these earlier Zr abundances have been revised and updated in this paper, as discussed in Section 3.   \citetalias{2012AJ....144...95Y} also noted that the trend of Zr with age was more moderate than that cited by \citetalias{2011ApJ...736..120M}.  However, as with La, the [Zr/Fe] abundances for the \citetalias{2011ApJ...736..120M} sample are consistent with the distribution of points shown in the bottom panel of Figure \ref{sfe_m_age}, and support the conclusion that there is no trend of [Zr/Fe] abundances with age among the open clusters from this larger sample.

\begin{figure*}
\epsscale{0.8}
\plotone{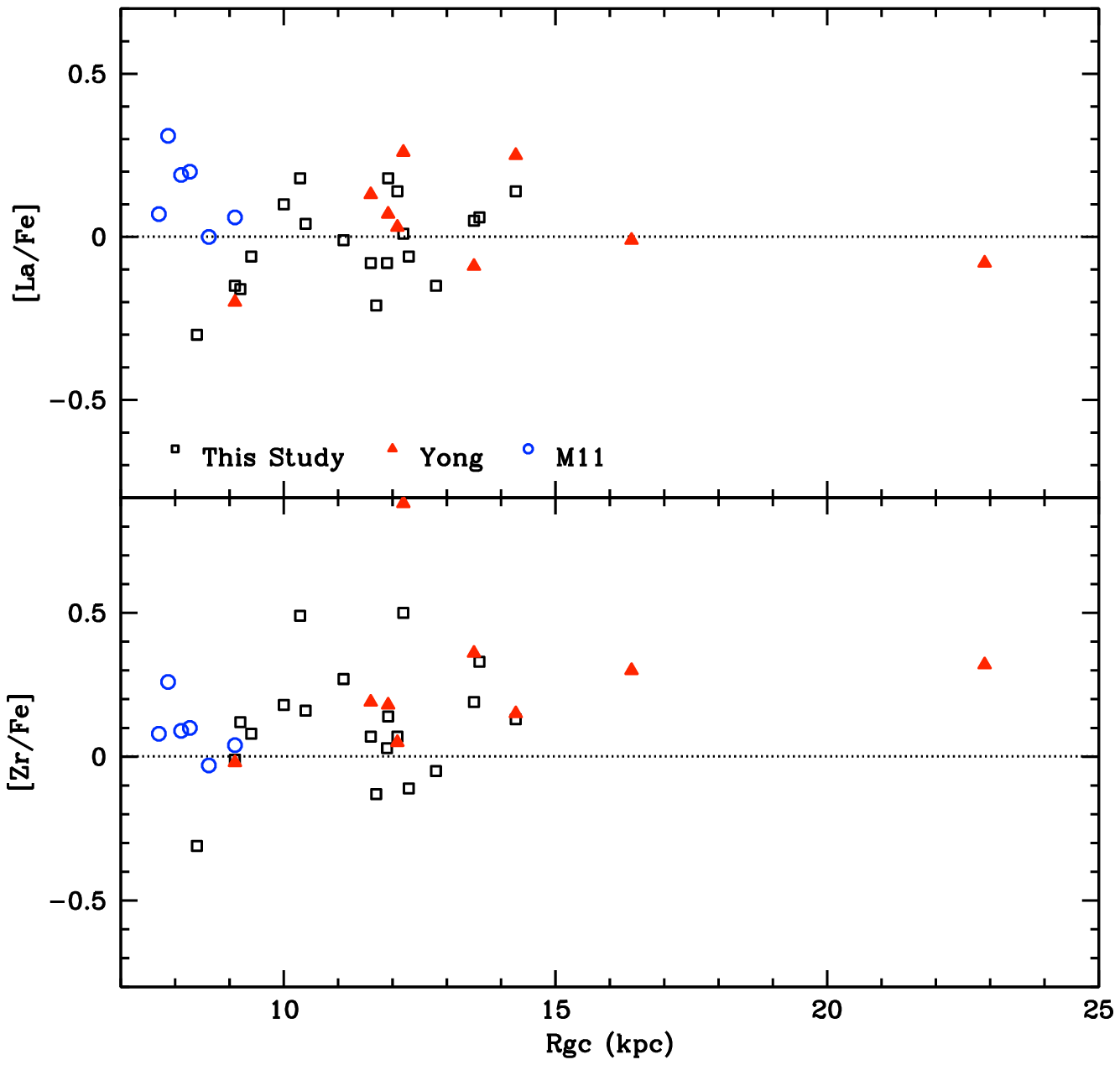}
\caption{[s/Fe] ratios as a function Galactocentric distance, symbols same as in Figure \ref{sfe_m_age}.}
\label{sfe_m_rgc}
\end{figure*}

Very few determinations of Eu exist for stars in open clusters, yet the behavior of this r-process element relative to the s-process elements we have considered thus far is especially interesting.  Figure \ref{eu_age} shows cluster [Eu/Fe] trends versus age for our sample.  Our data show a trend of {\it increasing} [Eu/Fe] with {\it increasing} cluster age but only at a marginally statistically significant level.  What is more notable in these figures is how the dispersion in cluster [Eu/Fe] ratio varies with age: for example, the dispersion in [Eu/Fe] among 4 Gyr old clusters spans nearly 0.6 dex, while the dispersion of the older and younger clusters is 0.3-0.4 dex.  Clearly, this could just be a result of the small size of our cluster sample, but it would be interesting to see if these dispersions hold in larger homogeneous cluster studies.  \citetalias{2012AJ....144...95Y} found no trend of [Eu/Fe] with age in their analysis, based on their results for 10 clusters plus four others from the literature.   Those results are consistent with our larger sample, as can be seen in the bottom panel of Figure \ref{eu_age}.

As mentioned in the introduction, \citet{2012A&A...542A..84D} found trends of increasing [s/Fe] with decreasing age for dwarf stars in the solar neighborhood, most notably for Ba, but also for Zr, Y and Ce.  Of these elements, only Y shows this anticorrelation across the full age range of their sample; the anticorrelation is present for the other elements only for the stars younger than the Sun (4.5 Gyr).  They argue, as \citetalias{2009ApJ...693L..31D} and \citetalias{2011ApJ...736..120M} do, that this indicates that the s-process yields of AGB stars has changed over time (as well as [Fe/H] of the AGB stars), and the fact that Ba shows the strongest trend with age implies that this change in yields affects the heavy s-process elements more than it does the light ones.  While we find the consistency between the results of \citet{2012A&A...542A..84D} and \citetalias{2009ApJ...693L..31D}/\citetalias{2011ApJ...736..120M} compelling, we would argue that the [La/Fe] trends seen in Figures \ref{sfe_age} and \ref{sfe_m_age} of this work may indicate that the source of an [s/Fe] abundance trend with age does not affect all heavy s-process elements equally.

Indeed, the picture becomes more complicated when other (larger) solar neighborhood dwarf star studies are considered.  We performed a regression analysis of [X/Fe] versus age for solar neighborhood (thin disk) dwarf stars from the studies of \citet{2003MNRAS.340..304R} and \citet{2005A&A...433..185B}, considering the full age range of their samples as well as stars 4.5 Gyr or younger.  For the \citet{2003MNRAS.340..304R} sample,  [Zr/Fe] showed a statistically significant trend with decreasing stellar age for both age ranges considered, while [Nd/Fe]\footnote{Technically, Nd is $\sim$50\%\ s-process and $\sim$50\%\ r-process \citep{2000ApJ...544..302B}.} showed a statistically significant trend with (increasing) age only when the full age range the sample was considered.  For the \citet{2005A&A...433..185B} sample, only [Ba/Fe] showed a statistically significant trend with (decreasing) age, for both age ranges considered.  None of the other s-process species studied by these two groups (Y for \citealt{2005A&A...433..185B}; Ba and Ce for \citealt{2003MNRAS.340..304R}) showed trends with age.  (The lack of a trend of [Ba/Fe] with decreasing cluster age for the \citet{2003MNRAS.340..304R} sample may be due to the fact that that sample contained very few stars with young ($<$1-2 Gyr) ages.  If the distribution of [Ba/Fe] ratios with age is non-linear, and in fact only becomes steeply anti-correlated with age for the youngest objects (e.g., the youngest clusters in \citetalias{2009ApJ...693L..31D}), then such a trend would not appear in a sample of older stars.)

\begin{figure*}
\epsscale{0.8}
\plotone{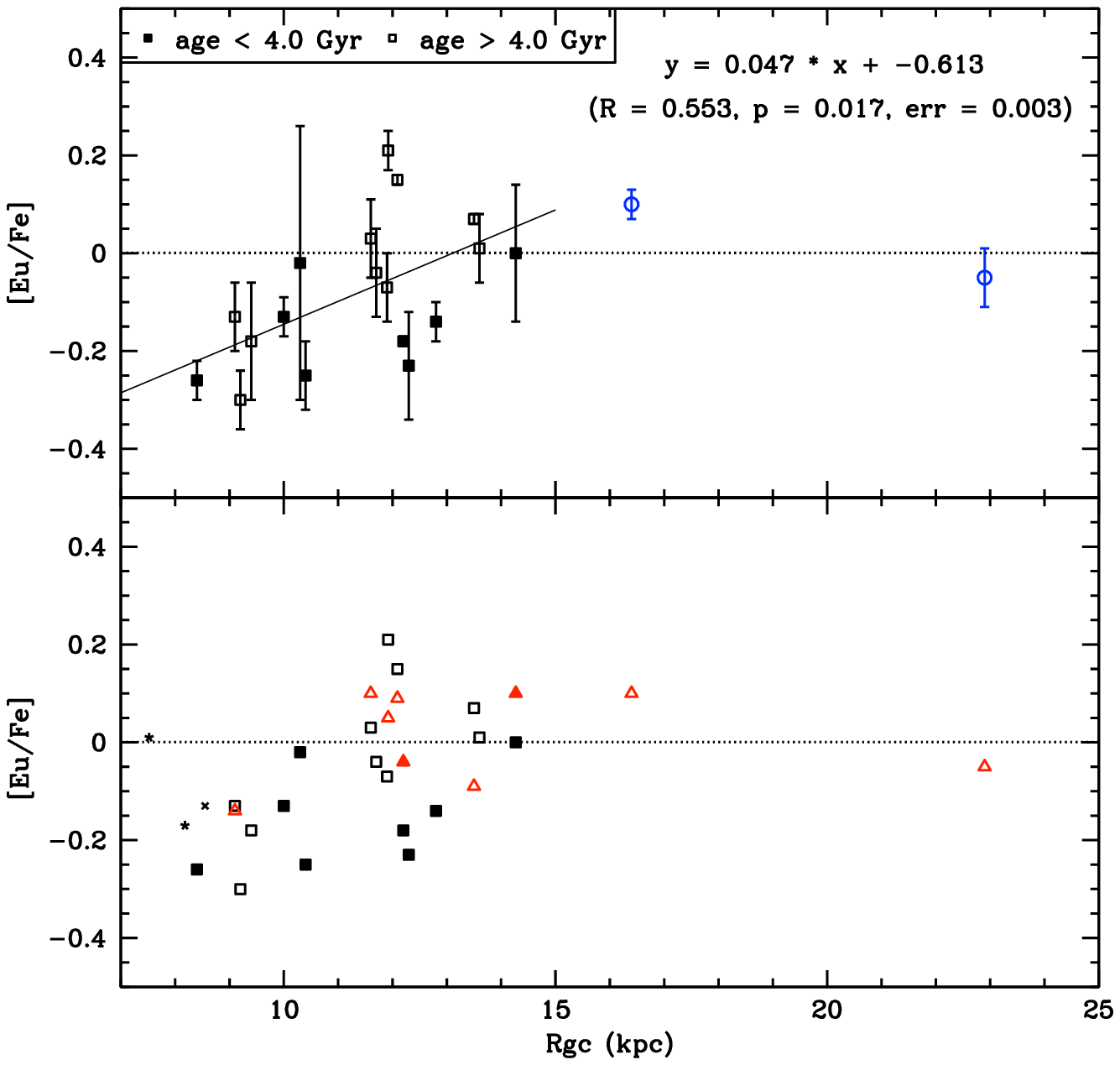}
\caption{Top panel: [Eu/Fe] ratios as a function Galactocentric distance.  Symbols same as in Figure \ref{sfe_rgc}.  Bottom panel: [Eu/Fe] ratios versus \rgc\ for our sample (squares), those of Yong et al.\ (2005, 2012; triangles) shifted to our abundance scale.  As in Figure \ref{sfe_rgc}, open symbols represent clusters 4 Gyr and older and filled symbols represent clusters younger than 4 Gyr.  Smaller points represent clusters from the literature (crosses: age $<$ 4 Gyr; asterisks: age $\geqq$ 4 Gyr).}
\label{eufe_rgc}
\end{figure*}

Regarding r-process abundance trends with age, \citet{2012A&A...542A..84D} did not determine Eu abundances for their sample.  However, they did report abundances for Sm, which is also representative of the r-process.  They also reported a trend of increasing [Sm/Fe] with increasing age for their sample, consistent with the Eu results for our open cluster sample.  The \citet{2003MNRAS.340..304R} sample also showed statistically significant trends of [Eu/Fe] with increasing stellar age, but the \citet{2005A&A...433..185B} sample did not.  

In Figure \ref{seu_age} we explore trends of [X/Eu] ratios versus age for Ba, La and Zr individually.  Trends of increasing [Ba/Eu] and [La/Eu] with decreasing cluster age can be seen, while that for [Zr/Eu] is less convincing.  These trends are consistent with the [s/Sm] versus age trends in \citet{2012A&A...542A..84D}, although the trends in their sample are especially pronounced for stars younger than the Sun.  We interpret this as evidence of the increasing importance/efficiency of s-process enrichment relative to the r-process in the Milky Way thin disk over time.

This analysis did not take into account uncertainties in the ages of the clusters, which can sometimes be significant.  We have chosen to use the cluster ages as determined in \citet{2004A&A...414..163S} which rest on a calibration of the morphological age indicator of \citet{1994AJ....108.1773J}.  While this method may not be as accurate for any given cluster as direct fitting of isochrones to cluster color magnitude diagrams, it has the advantage of placing all the cluster ages on a uniform relative scale.  The typical age uncertainty of the \citeauthor{2004A&A...414..163S} calibration is 15\%, which for our clusters is less than 1 Gyr.   To give an indication of the impact of age uncertainties on the results, we varied the age of the oldest cluster, Be 17, by its uncertainty of 2.77 Gyr as given by the \citeauthor{2004A&A...414..163S} calilbration.   This estimate of the uncertainty in age offers an extreme case, both in terms of the  \citeauthor{2004A&A...414..163S} calibration and direct age determinations (e.g., \citealt{2006MNRAS.368.1971B} found an age of 8.5 - 9 Gyr for Be 17 from isochrone fits).  Making Be 17's age younger or older both served to slightly decrease the statistical significance of the trends shown in Figure 3, e.g., p=0.013 and 0.017 for [Ba/Fe].    Coupled with the leverage Be 17's large age gives to the fitting of element abundance trends with age, this example provides an indication of the robustness of such trends to age errors.  Adopting the more typical 15\% age uncertainty of the \citeauthor{2004A&A...414..163S} calibration would have significantly less impact on any of these correlations.

Lastly, we also performed a regression analysis of our sample's [X/Fe] ratios as a function of log age, such as seen in Figure \ref{ba_age}, to assess how sensitive the statistical significance of the trends were to how age was represented.  We found that, in general, the p-values increased for each element considered here, decreasing the statistical significance of each trend.  For example, the p-value of [Ba/Fe] versus log age was 0.041, as opposed to 0.010 as seen in Figure \ref{sfe_age}.  However, the opposite is found in an analysis of the \citetalias{2009ApJ...693L..31D} sample: the p-value of their cluster [Ba/Fe] ratios versus log age was 0.0003, compared to p = 0.011 found versus age.  This is no doubt due to the much larger age range of the \citetalias{2009ApJ...693L..31D} sample relative to ours (the youngest being 35 Myr old, as opposed to our ~700 Myr).  These changes  in p-value do not change the primarily conclusions described above. 

\begin{figure*}
\epsscale{0.8}
\plotone{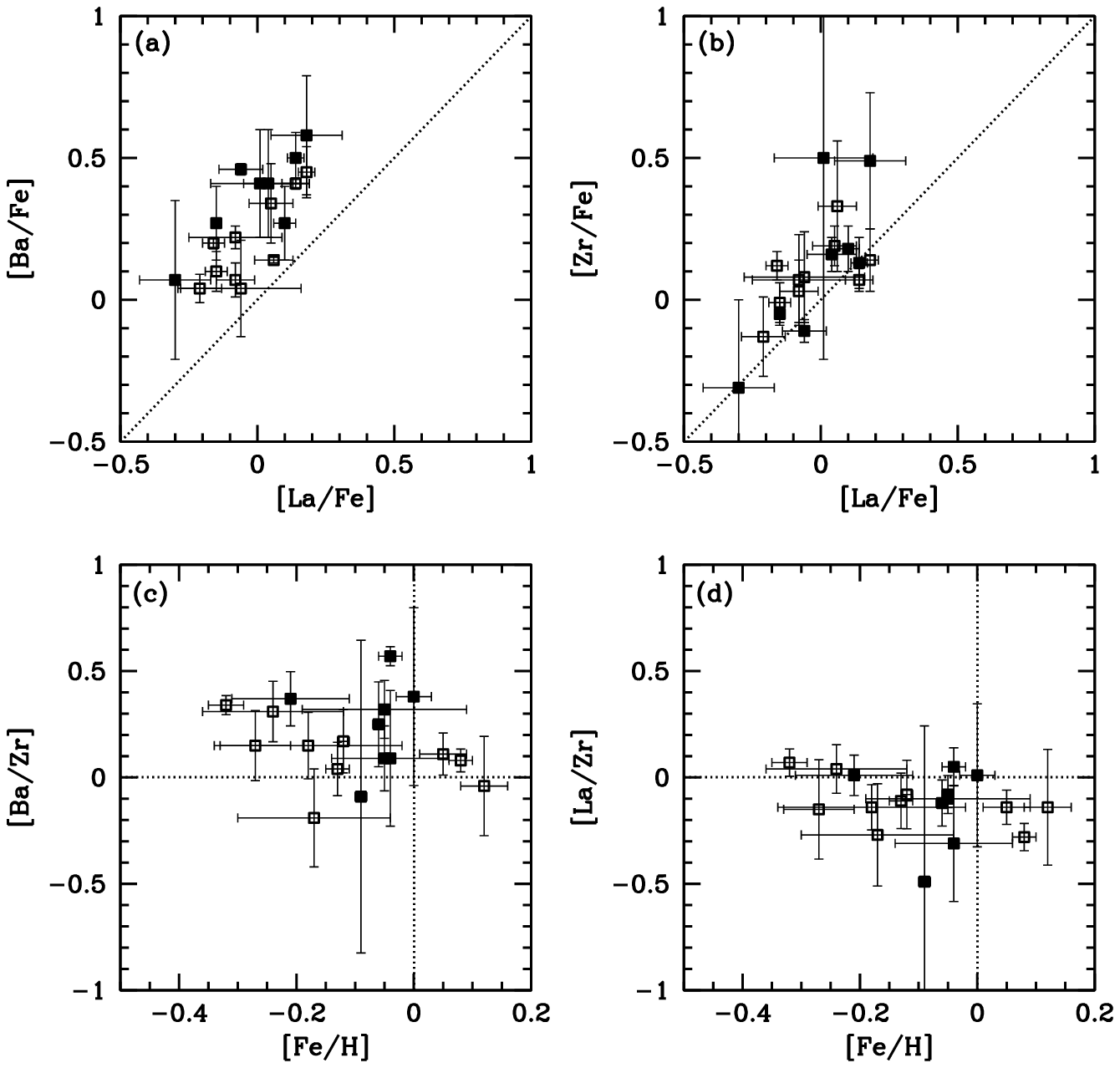}
\caption{Top panels: [Ba/Fe] (a) and [Zr/Fe] (b) versus [La/Fe] for our sample, with dotted lines indicating the 1-to-1 relation.  Bottom panels: [Ba/Zr] (c) and [La/Zr] (d) versus [Fe/H] for our sample.  Symbols same as in Figure \ref{sfe_rgc}.}
\label{s_vs_la_fe}
\end{figure*}

\subsection{Cluster n-capture abundance trends with R$_{gc}$}

The open clusters have often been used as tracers of the elemental abundance distributions in the Milky Way disk.  In terms of the traditional characteristics used to categorize disk populations, the properties of the open clusters show overall consistency with those of the thin disk (Friel 1995).  Their kinematics, and in particular the systemic rotational properties of the open cluster population, are very similar to those of the thin disk field stars and other disk tracers \citep{1987A&A...176...34H,1995AJ....109.1706S}.  Cluster ages, spatial distribution, and concentration to the Galactic plane are thin disk-like.  As demonstrated by many papers (e.g., \citealt{2011AJ....142...59J,2009A&A...494...95M}, the elemental abundance ratios of open clusters follow the trends shown by field stars of the thin disk.   Although there may be a few open clusters whose individual properties distinguish themselves in interesting ways (e.g. NGC 6791 - \citealt{2011ApJ...733L...1P,2012ApJ...756L..40G}), the evidence for the open cluster population as a whole is that it provides an important and useful sample with which to study the evolution of the thin disk properties.

It is established by now that open clusters (and other disk populations) show that the thin disk's metallicity gradient, which decreases roughly with a slope of $-$0.07 dex kpc$^{-1}$ through the solar neighborhood (e.g., \citealt{2011A&A...535A..30C}), flattens out at [Fe/H] $\sim$$-$0.3 around \rgc $\sim$13 kpc.  Cluster [X/Fe] ratios are essentially flat across the full \rgc\ range spanned by open clusters for the light, $\alpha$, and Fe-peak elements, indicating that they all follow Fe fairly closely.  It is therefore interesting, especially in light of the trends of at least some of them with cluster age, to investigate the trends of n-capture element abundances with \rgc.  To this end, \citetalias{2011ApJ...736..120M} and Yong et al.\ (2005, 2012) have all investigated n-capture abundances with \rgc.  \citetalias{2011ApJ...736..120M} remarked that the s-process elements they studied have distributions similar to the other element groups studied (i.e., $\alpha$, Fe-peak) -- that is, that cluster [s/Fe] is essentially independent of \rgc.  That said, they note that the trends with \rgc\ for young and old clusters are different, as expected based on the abundance trends with age found in their sample (see, e.g., their Figures 6 and 7  for Ce and Y abundances).  

\citetalias{2012AJ....144...95Y} investigated radial abundance trends in clusters with a focus on distinguishing behavior in the solar neighborhood from that in the outer disk (taken to be Rgc $>$ 13 kpc).  For the neutron-capture elements they find that [Zr/Fe] shows no trend with \rgc, but  for Ba, La, and Eu, there is a suggestion that the outer disk clusters follow a different gradient from that of the local sample of clusters.  In particular, for La and Eu, the outer disk clusters appear enhanced relative to those inside $\sim$ 10 - 13 kpc.

\begin{figure*}
\epsscale{0.8}
\plotone{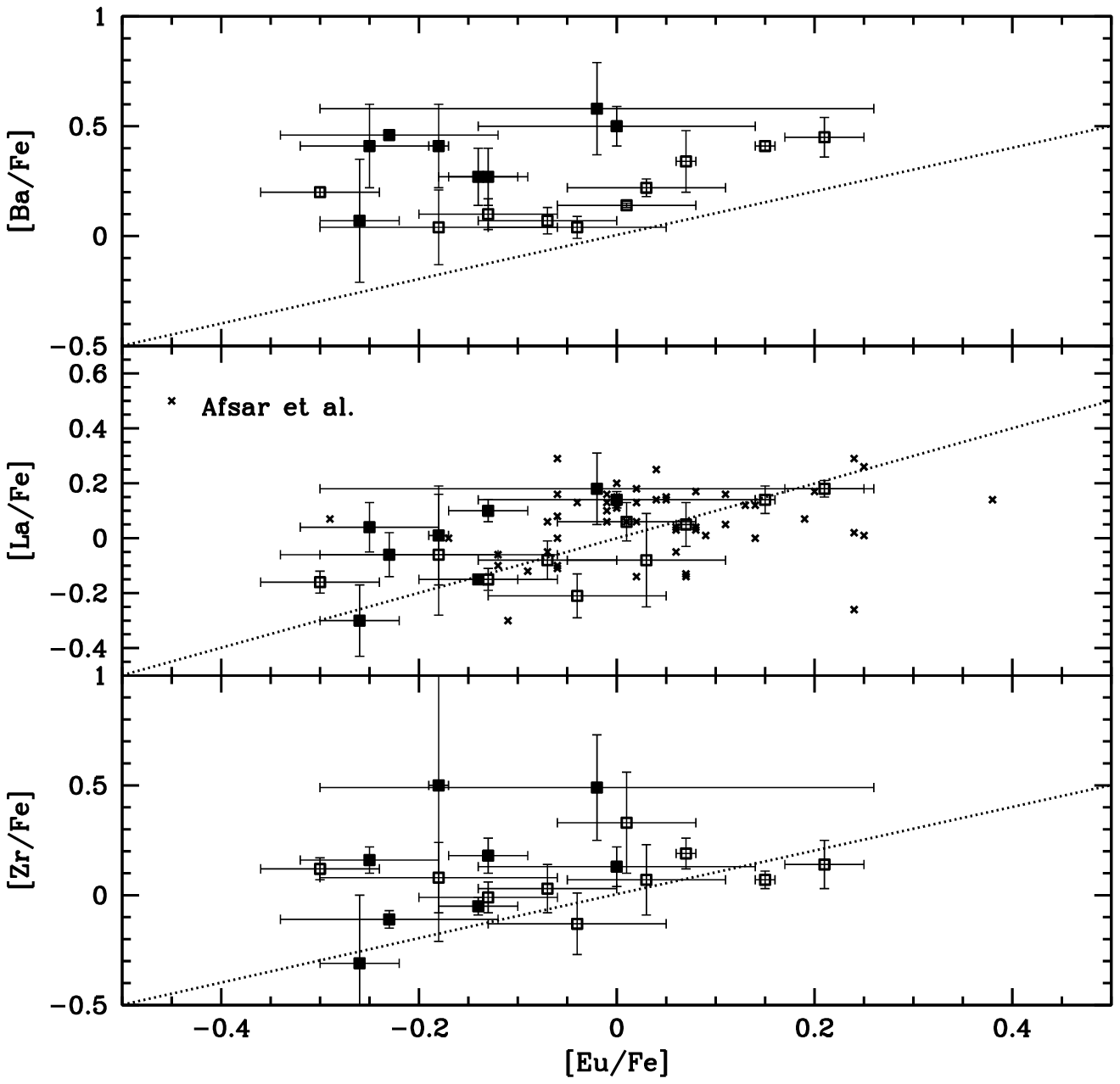}
\caption{[X/Fe] versus [Eu/Fe] for Ba (top), La (middle) and Zr (bottom). Symbols are the same as in Figure \ref{sfe_rgc}, along with field red horizontal branch stars  from ~\citet{2012AJ....144...20A} (black crosses) in the middle panels.  Dotted lines represent the 1-to-1 relation.}
\label{sfe_vs_eu}
\end{figure*}

We present [Ba/Fe], [La/Fe] and [Zr/Fe] ratios as a function of cluster \rgc\ for our sample in Figure \ref{sfe_rgc}.  In each panel, we distinguish older (4 Gyr and greater) and younger (younger than 4 Gyr) clusters by open and filled symbols, respectively, though the details of regression analyses given in each panel is for the entire cluster sample.  Given the narrow \rgc\ range of the bulk of our cluster sample (\rgc$\sim$8-14 kpc), we do not draw conclusions about general abundance distributions in the disk as a whole.  Rather, we present a linear regression analysis of our original cluster sample alone (with \rgc$<$14 kpc) and include the results in each panel for comparison to the gradients presented by \citetalias{2012AJ....144...95Y} for what they term the ``inner disk" (\rgc$<$13 kpc; their Figure 24).  Generally, the gradients we find for [La/Fe] and [Zr/Fe] are larger than those found by \citetalias{2012AJ....144...95Y}, while our slope for [Ba/Fe] is shallower than theirs.  None of these slopes are statistically significant.  Figure \ref{sfe_m_rgc} shows [La/Fe] and [Zr/Fe] for our sample along with that of \citetalias{2012AJ....144...95Y} and \citetalias{2011ApJ...736..120M} as a function of \rgc.  Again, the distributions of all three samples are consistent with one another.  

Figure \ref{eufe_rgc} shows [Eu/Fe] versus \rgc, again with the clusters divided into younger (filled symbols) and older (open symbols) age bins.  
\citetalias{2012AJ....144...95Y} reported a possible trend of enhanced [Eu/Fe] in the outer disk.  However, with their [Eu/Fe] ratios adjusted to our abundance scale, the four clusters beyond \rgc$\sim$13 kpc in their sample have abundance ratios consistent with the inner disk clusters (bottom panel of Figure \ref{eufe_rgc}).
Therefore, in comparison to the conclusions from \citetalias{2012AJ....144...95Y}, we note that the systematic differences between our abundance scales for [La/Fe] ($-$0.29 dex) and [Eu/Fe] ($-$0.16 dex), when taken into account, would serve largely to remove the differences they see between the inner and the outer disk abundances, yielding trends consistent with those seen in Figures \ref{sfe_rgc} and \ref{eufe_rgc}.

That said, the younger and older populations of clusters appear distinct in [Eu/Fe] versus \rgc\ (Figure \ref{eufe_rgc}).  
The younger clusters (filled symbols) do not exhibit statistically significant trend with \rgc, but the trend with \rgc\ for the older clusters (open symbols) is marginally significant (p=0.017).  We found, however, that these trends are largely driven by the radial variation of cluster [Fe/H] rather than of [Eu/H], indicative of a possible flattening of the thin disk abundance gradient over time.  It would be interesting to explore possible variations in the radial distributions of younger and older clusters in larger, homogeneous cluster samples that better probe the full \rgc\ and age range of the Galactic disk.

\subsection{Other abundance ratio trends}
As discussed in earlier sections, barium and lanthanum are both heavy s-process elements, so it seems reasonable to expect their abundances to scale together.  Figure \ref{s_vs_la_fe} shows [Ba/Fe] versus [La/Fe] (top panels)  for our cluster sample.    Here again, we distinguish younger clusters ($<$4 Gyr) from older clusters ($>$4 Gyr) with filled and open symbols, respectively.  It can be seen from the top left panel that [Ba/Fe] and [La/Fe] do appear to scale together, though with a constant offset from a 1-to-1 correlation.  [Zr/Fe] also seems to be well correlated with [La/Fe] for most clusters in our sample as well (top right panel).  Also, younger clusters in our sample are slightly shifted to higher [Ba/Fe] ratios from the older clusters as a result of the [Ba/Fe] correlation with age we have already discussed, but with much overlap (especially when errorbars are taken into account).

\begin{figure*}
\epsscale{0.8}
\plotone{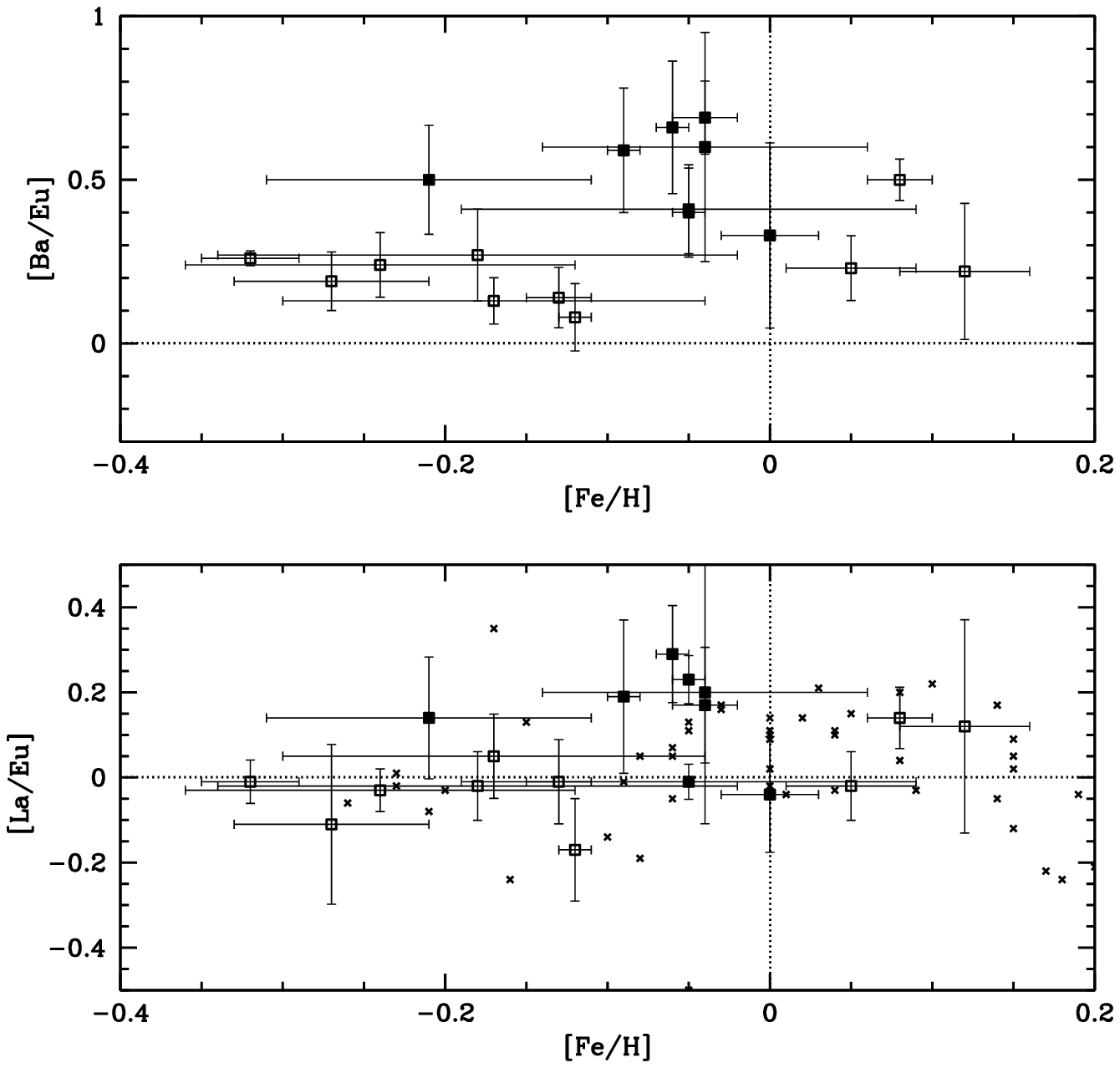}
\caption{[Ba/Eu] (top panels) and [La/Eu] (bottom panels) versus [Fe/H] for our sample.  Symbols are the same as in Figure \ref{sfe_rgc}, and again we include the RHB stars from \cite{2012AJ....144...20A} as small crosses.}
\label{rs_vs_fe}
\end{figure*}

As Zr is a light s-process element, the ratios of [Ba/Zr] and [La/Zr] provide information about the efficiency of s-process element production in the AGB stars that enriched the gas from which clusters in the disk formed\footnote{As mentioned previously, as a light s-process element, Zr can also be produced via the weak s-process in massive stars (e.g., \citealt{1985A&A...151..205B}).  However, as mentioned in \citet{2012ApJ...747...53M}, the estimated weak s-process contribution to the solar system light element abundance distribution is small \citep{2009PASA...26..153S}.}.  The bottom panels of Figure \ref{s_vs_la_fe} shows the ratios of Ba and La to Zr as a function of cluster [Fe/H] for our sample with the clusters distinguished by age.  Neither age group shows a statistically significant trend of [X/Zr] with [Fe/H].
Indeed, for the cluster sample as a whole, these [hs/ls] versus [Fe/H] trends are rather flat, similar to those found for thin disk dwarf stars in \citeauthor{2003MNRAS.340..304R} (2003; their Figure 14).  We note that plots of [Ce/Y] versus [Fe/H] for the \citetalias{2011ApJ...736..120M} sample also show a flat distribution with metallicity.  \citeauthor{2003MNRAS.340..304R} say, ``The similarity of the slopes of [X/Fe] versus [Fe/H] for light (Sr, Y, Zr) and heavy (Ba, Ce) elements may reflect either unchanging relative contributions from AGB stars or a change in these contributions that is offset by a change in the weak s-process contributions from the massive stars.  On the assumption that the AGB stars are the controlling influence, the unchanging abundance ratio of heavy to light elements indicates that the exposures to neutrons in the s-process site is essentially independent of the metallicity of the contributing AGB stars."  If this interpretation is true, it implies that s-process yields of AGB stars do not change as a function of [Fe/H] in the metallicity range spanned by our sample, but may still vary as a function of AGB star age (i.e., mass), as mentioned in \citet{2012A&A...542A..84D}.  This interpretation is consistent with the model proposed by \citet{2012ApJ...747...53M}, but again the situation is confused by the  uncertainties in the abundance trends with age seen for the heavy and light s-process elements among the various studies.

Figure \ref{sfe_vs_eu} plots [Ba/Fe], [La/Fe] and [Zr/Fe] versus [Eu/Fe] for clusters in this study.  In addition, we also show [La/Fe] versus [Eu/Fe] for red horizontal branch stars from the study of \citet{2012AJ....144...20A}.  Plots of s-process elements versus r-process elements reveals the relative contributions of the s- and r-processes to the chemical enrichment of the Milky Way disk.  Another presentation of [s/r] ratios relative to [Fe/H] is given in Figure \ref{rs_vs_fe}.  As can be seen in the bottom panel, the older clusters in our sample (open squares) have scaled solar [La/Eu] ratios, except for the two clusters with super-solar metallicity (NGC 7142 and NGC 188).  The younger clusters, on the other hand, generally show enhanced [La/Eu] ratios.  A similar pattern is shown for [Ba/Eu], too, shifted relative to that for La (recall the top left panel of Figure \ref{s_vs_la_fe}).  The distribution of [La/Eu] versus [Fe/H] is similar to that shown by the RHB stars of \citeauthor{2012AJ....144...20A} as well, indicating that metal-rich RHB stars are comparable to clusters as tracers of thin disk characteristics, as \citeauthor{2012AJ....144...20A} conclude.  Taken altogether, these figures as well as Figure \ref{seu_age} show the increasing dominance of the s-process to the chemical evolution of the galactic thin disk.

\section{Summary and Conclusions}
The context of this work has been the recently-discovered trends of open cluster neutron capture element abundances with age and \rgc.  \citetalias{2009ApJ...693L..31D} and \citetalias{2011ApJ...736..120M} have identified increasing [s/Fe] with decreasing age for open clusters that affects both heavy and light s-process elements; \citetalias{2012AJ....144...95Y} found such a trend only for Ba.  We have also remarked on the analysis of solar neighborhood dwarf stars by \citet{2012A&A...542A..84D}, who also reported trends of increasing [s/Fe] ratios with age, especially for stars younger than 4.5 Gyr, though we noted the contradictory findings of other solar neighborhood field star studies \citep{2003MNRAS.340..304R,2005A&A...433..185B}.  

Here, we have presented an analysis of neutron capture element abundances in a sample of 19 open clusters that is completely independent of the open cluster studies described above.  In this context, we summarize the results of this work as the following:
 
\begin{itemize}
\item We have found a statistically significant trend of increasing cluster mean [Ba/Fe] with decreasing cluster age, in agreement with \citetalias{2009ApJ...693L..31D}, \citetalias{2012AJ....144...95Y} and \citet{2012A&A...542A..84D}.
\item In contrast, [La/Fe] and [Zr/Fe] abundances for our sample do not show trends with age.  The distribution of [La/Fe] and [Zr/Fe] abundances of the \citetalias{2011ApJ...736..120M} and \citetalias{2012AJ....144...95Y} samples are not inconsistent with our results.
\item We report a marginally significant trend of increasing cluster mean [Eu/Fe] with increasing cluster age for our sample, while trends of [Ba/Eu] and [La/Eu] versus age are statistically significant.  This indicates the increase in dominance of s-process enrichment to the thin disk over time.
\item Investigations of [s/Fe] versus \rgc\ for our cluster sample found no statistically significant trends, bearing in mind the limited \rgc\ range of our sample ($\sim$9-13 kpc).  That said, clusters older than 4 Gyr do show a marginally statistically significant increase in [Eu/Fe] versus \rgc\ that the younger clusters do not.  This may be an indication of the flattening of disk abundance gradients with time \citep{2002AJ....124.2693F,2011AJ....142...59J}.
\item The ratio of heavy to light s-process element abundances ([hs/ls]) versus [Fe/H] for our cluster sample is flat, consistent with the results of \citetalias{2011ApJ...736..120M} and \citet{2003MNRAS.340..304R} for thin disk dwarfs.  One interpretation of this result is that the efficiency of the s-process in AGB stars is independent of their metallicity in this metallicity range, as suggested by \citet{2003MNRAS.340..304R}.
\item The younger clusters in our sample show enhanced [s/r] ratios compared to the older open clusters, as one might expect based on the trends of individual elements with age discussed above.  There is no trend of [s/r] ratio with [Fe/H], again indicating that metallicity does not seem to play a crucial role in evolution of neutron capture elements in open clusters (or more explicitly, in open clusters spanning the metallicity range of our sample: [Fe/H]$\sim$ $-$0.30 to $+$0.10).  The clusters' [s/r] ratio is also seen to be consistent with other disk populations (RHB stars).
\end{itemize}

How does one resolve the inconsistencies between the results of this study, \citetalias{2009ApJ...693L..31D}/\citetalias{2011ApJ...736..120M} and \citetalias{2012AJ....144...95Y}, as well as the inconsistencies in the behaviors of the heavy and light s-process elements?  Are these all due to errors or a too-large dispersion in, e.g., our Zr and La abundances that mask trends that are easily seen for Ba?  Given the strength of the Ba lines and the sensitivity of the EW measurements to continuum placement and/or blending, as well as their sensitivity to uncertainties in microturbulent velocity, we consider our La measurements to be more reliable.  It is possible that the lack of clusters younger than 700 Myr in our sample and in \citetalias{2012AJ....144...95Y} may limit the detectability of trends with age if only the youngest objects have large enhancements in [s/Fe].  It is difficult to disentangle the impacts of different age ranges and systematic abundance differences on the conclusions found by different studies.  
Clearly, it imperative to expand the analysis to include other neutron capture species (both s- and r-process) in not only this sample, but in a larger open cluster sample.  We look forward to the Gaia-ESO Survey of stars in many dozens of open clusters that will facilitate the measure of many n-capture species in clusters spanning a wide range in age \citep{2012Msngr.147...25G}.  We hope that the existence and strength of any abundance trends with age so far seen (or not) can be verified in a larger, homogeneous study of open cluster chemical abundances.

\acknowledgements 

We gratefully acknowledge the referee for a thoughtful report that improved the presentation of this work.  We thank David Yong for sharing reduced spectra of many stars used in this analysis.  HRJ thanks Christian Johnson for sharing line lists and for many helpful discussions regarding spectrum synthesis with MOOG, as well as Nick Sterling for many conversations about the s-process.  
This research has made use of NASA's Astrophysics Data System Bibliographic Services and the Vienna Atomic Line Database.
This work has been carried out with support from the National Science Foundation through the NSF Astronomy and Astrophysics Postdoctoral Fellowship under award AST-0901919.

\bibliographystyle{apj}
\bibliography{mybib}

\end{document}